\documentclass[1p,number,times]{elsarticle}
\usepackage[usenames]{color} 
\usepackage{ulem} 
\usepackage{graphicx}
\usepackage{amssymb} 
\usepackage{amsmath}

\begin{document}
 

\title{Solving Partial Differential Equations Numerically
on Manifolds with Arbitrary Spatial Topologies}

\author[caltech]{Lee Lindblom} 
\author[caltech]{B\'ela Szil\'agyi}

\address[caltech]{Theoretical Astrophysics 350-17, California Institute of
Technology, Pasadena, CA 91125}

\date{\today}
 
\begin{abstract} A multi-cube method is developed for solving systems of
elliptic and hyperbolic partial differential equations numerically on
manifolds with arbitrary spatial topologies.  It is shown that any
three-dimensional manifold can be represented as a set of
non-overlapping cubic regions, plus a set of maps to identify the
faces of adjoining regions.  The differential structure on these
manifolds is fixed by specifying a smooth reference metric tensor.
Matching conditions that ensure the appropriate levels of continuity
and differentiability across region boundaries are developed for
arbitrary tensor fields.  Standard numerical methods are then used to
solve the equations with the appropriate boundary conditions, which
are determined from these inter-region matching conditions.  Numerical
examples are presented which use pseudo-spectral methods to solve
simple elliptic equations on multi-cube representations of manifolds
with the topologies $T^3$, $S^2\times S^1$ and $S^3$.  Examples are
also presented of numerical solutions of simple hyperbolic equations
on multi-cube manifolds with the topologies $R\times T^3$, $R\times
S^2\times S^1$ and $R\times S^3$.
\end{abstract}

\begin{keyword}
topological manifolds \sep numerical methods \sep partial differential
equations
\end{keyword} 
 
\maketitle
\section{Introduction}
\label{s:Introduction}

The need to solve partial differential equations on manifolds having
non-trivial spatial topologies arises in many areas of physical
science: from models of wormholes or the global structure of the
universe in general relativity theory to global circulation models of
the earth's atmosphere in meteorology and climatology.  This paper
develops practical methods for solving a variety of partial
differential equations on manifolds having arbitrary spatial
topologies.  Every $n$-dimensional manifold (by definition) can be
mapped locally into a portion of $n$-dimensional Euclidean space,
$R^{\,n}$.  A number of different numerical methods are capable of
solving partial differential equations locally on open subsets of
$R^{\,n}$.  The topological structure of a manifold, however, affects
the global solutions to partial differential equations in profound
ways.  This paper develops methods for fitting together local
solutions, obtained from standard numerical methods, to form the
desired global solutions on manifolds with arbitrary topologies.  The
discussion here focuses on solving elliptic systems of equations on
three-dimensional manifolds $\Sigma$ with arbitrary topologies, and
also hyperbolic systems of equations on four-dimensional manifolds
with topologies $R\times\Sigma$.

Solving partial differential equations numerically on manifolds with
arbitrary topologies requires the creation of computational
infrastructures (beyond those needed to solve the equations
numerically on open subsets of $R^{\,n}$) that meet two basic
requirements.  The first requirement is that the manifold must be
represented in a way that allows the points in the manifold, and the
values of scalar and tensor fields defined at those points, to be
referenced efficiently in a way that respects the underlying
topological structure of the manifold.  The second requirement is to
create a way to specify the global differential structure of the
manifold, i.e. the computational method must provide a way of
representing globally continuous and differentiable scalar and tensor
fields on these manifolds.  The goal here is to develop practical
methods that can be used on arbitrary manifolds by a wide range of
different numerical methods.

The first requirement is to find a systematic way of representing
manifolds with arbitrary topologies.  Every $n$-dimensional manifold
can be mapped locally into a portion of $n$-dimensional Euclidean
space $R^{\,n}$.  For computational efficiency (and to avoid certain
types of numerical instabilities) each manifold is represented here by
a collection of non-overlapping $n$-dimensional cubes which cover the
manifold, plus a set of maps that identify the faces of adjoining
$n$-cubes.  This decomposition is analogous to representing a manifold
as a collection of non-intersecting $n$-simplexes (i.e., triangles for
$n=2$ and tetrahedrons for $n=3$) that cover the manifold, plus maps
that identify neighboring faces.  Many numerical methods (including
the pseudo-spectral methods used to produce illustrative examples for
this paper) are easier to use in computational domains based on
$n$-cubes rather than $n$-simplexes.  Points in each of the $n$-cube
regions are identified by local Cartesian coordinates, and these
coordinates are used to represent the solutions to the differential
equations in each $n$-cube.  This type of representation has been used
for some time in numerical methods for solving partial differential
equations on a two-sphere~\cite{Ronchi1996, Taylor1997, Dennis2003},
and also in three-dimensional manifolds that are subsets of
$R^3$~\cite{Thornburg2000,Thornburg2004a,Lehner:2005bz, 
Schnetter:2006pg, Pazos:2009vb,Pollney:2009yz,Korobkin2011}.  Those
ideas are generalized in Sec.~\ref{s:TopologicalStructure}, and it is
shown that these generalizations can be applied to two-dimensional or
three-dimensional manifolds having arbitrary topologies.  Examples of
these multi-cube representations are given in
\ref{s:ExampleRepresentations} for the three-dimensional manifolds
with the topologies $T^3$, $S^2\times S^1$, and $S^3$.

The second requirement is to develop a method of representing (at
least in the continuum limit) continuous and differentiable tensor
fields on the multi-cube representations of manifolds developed in
Sec.~\ref{s:TopologicalStructure}.  Representing tensor fields within
each of the $n$-cube regions is straightforward: their components can
be expressed in the tensor bases associated with the local Cartesian
coordinates.  These tensor components are functions of those
coordinates, and their continuity (or differentiability) determines
the continuity (or differentiability) of the tensor field itself.  In
general, however, the coordinate tensor bases associated with
different $n$-cube regions are not even continuous (and can not be
made continuous globally) across the interfaces that join them.  The
problem of defining the continuity and differentiability of tensor
fields across $n$-cube interfaces is therefore non-trivial.  The
method introduced here makes use of a smooth reference metric
tensor. This reference metric must be supplied (along with the
collection of $n$-cube regions and the associated interface maps) as
part of the specification of a particular manifold.  This metric is
used to construct geometrical normal vectors at each interface, and
these normals are used to construct the Jacobian matrices that map
vectors (and tensors) across interfaces.  The differentiability of
tensors across the $n$-cube interfaces is defined in terms of the
continuity of the covariant derivatives of those tensors, using the
covariant derivative associated with the reference metric.  The
details of these continuity and differentiability conditions are given
in Sec.~\ref{s:DifferentialStructure}.  Examples of reference metrics
which can be used to implement these continuity and differentiability
conditions are given in \ref{s:ExampleRepresentations} for the
three-dimensional manifolds with the topologies $T^3$, $S^2\times
S^1$, and $S^3$.

Systems of differential equations can be solved numerically on
multi-cube representations of manifolds by fitting together local
solutions from each $n$-cube region.  The appropriate local solutions
are determined in each region by applying the correct boundary
conditions on the $n$-cube faces.  The appropriate boundary conditions
are the ones that enforce the needed level of continuity and
differentiability of the global solution at the region boundaries.
These boundary conditions are developed in
Sec.~\ref{s:BoundaryConditions} for second-order strongly elliptic
systems, and also for first-order symmetric hyperbolic systems of
equations.  These boundary conditions select the unique local solution
in a particular $n$-cube that equals the desired global solution in
that region.  The collection of local solutions to the equations
constructed in this way provides the desired global solution.

The multi-cube method of solving systems of partial differential
equations numerically on manifolds with non-trivial topologies is
illustrated here by solving a series of test problems in
Secs.~\ref{s:TestsEllipticEquations} and
\ref{s:TestsHyperbolicEquations}.  Simple second-order elliptic
equations, and first-order symmetric hyperbolic equations, are solved
numerically on manifolds with spatial topologies $T^3$, $S^2\times
S^1$, and $S^3$.  These tests use pseudo-spectral methods to produce
local solutions on each cubic region. The results are shown to
converge exponentially (in an $L^2$ norm) to the exact global
solutions (which are known analytically for these test problems) as
the number of grid points used for the solution is increased.


\section{Building Multi-Cube Manifolds}
\label{s:TopologicalStructure}

This section describes how $n$-dimensional manifolds can be
represented using the multi-cube method.  The idea is quite simple:
$n$-dimensional multi-cube representations of manifolds consist of a
set of non-overlapping $n$-cubes that cover the manifold, plus a set
of maps that identify the boundary faces of neighboring cubes.  An
argument is presented in Sec.~\ref{s:ExistenceOfMultiCubes} that all
two-dimensional and all three-dimensional manifolds (with arbitrary
topologies) can be represented in this way.  A large class (but not
all) higher-dimensional manifolds can also be represented using this
multi-cube method.  The multi-cube method provides a way of
representing manifolds that facilitates the design of computational
tools for solving partial differential equations on them.  A simple
infrastructure is introduced in
Sec.~\ref{s:InfrastructureForMultiCubes} for systematically building,
referencing and identifying the faces of the needed sets of $n$-cubes
in these manifolds.  These $n$-cube regions are joined together to
form the desired topological manifold using maps that identify points
on the faces of neighboring $n$-cubes.  A simple framework for
building and referencing these maps is presented.  Only a small number
of topologically distinct maps are needed for the case of
three-dimensional manifolds (the main focus of this paper), and all of
those maps are given explicitly.

\subsection{Existence of Multi-Cube Representations}
\label{s:ExistenceOfMultiCubes}

\begin{figure}[ht]
\begin{picture}(0,130)(0,75)
\put(25,80){
\includegraphics[width=2.in]{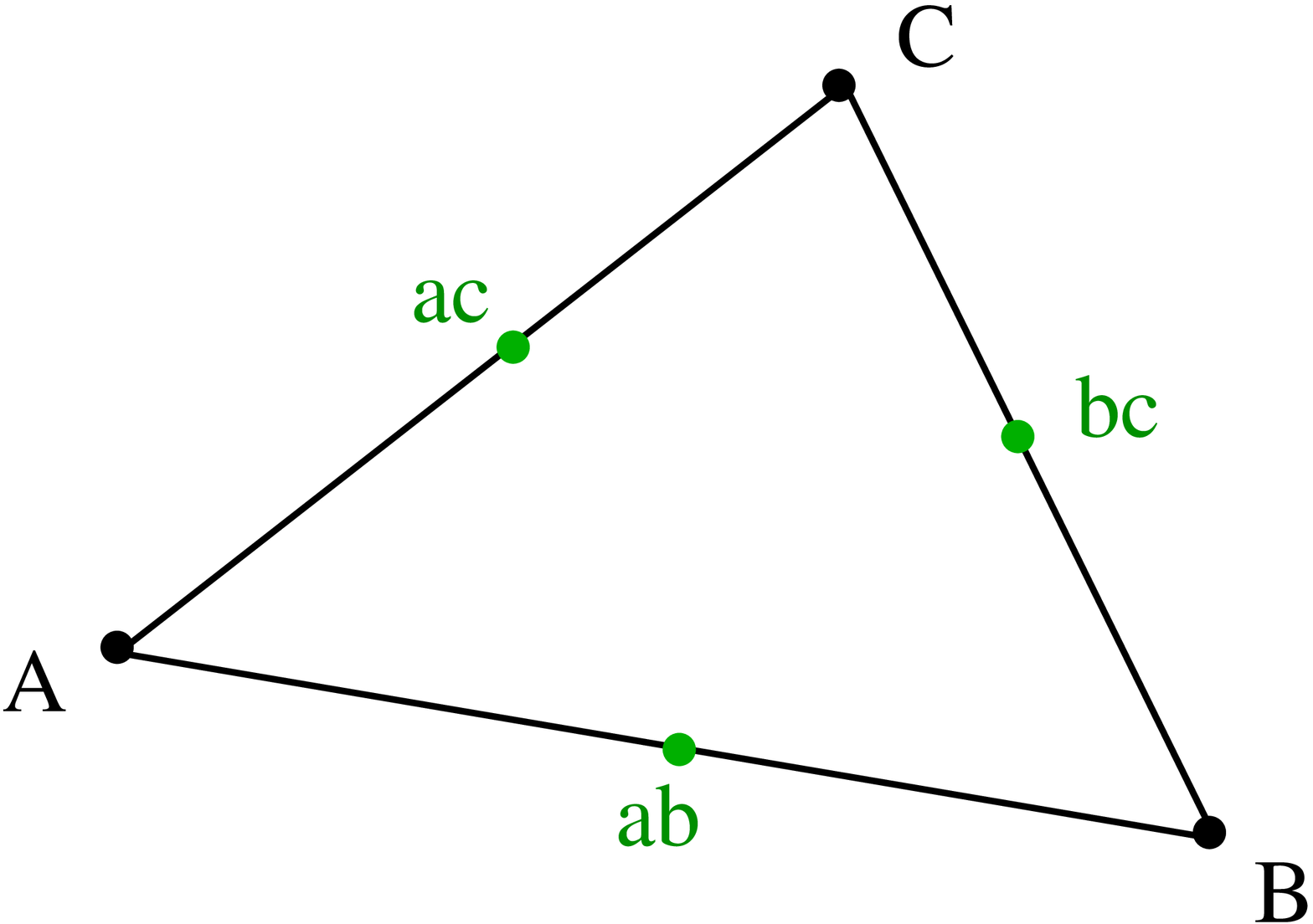}
} 
\put(200,80){
\includegraphics[width=2.in]{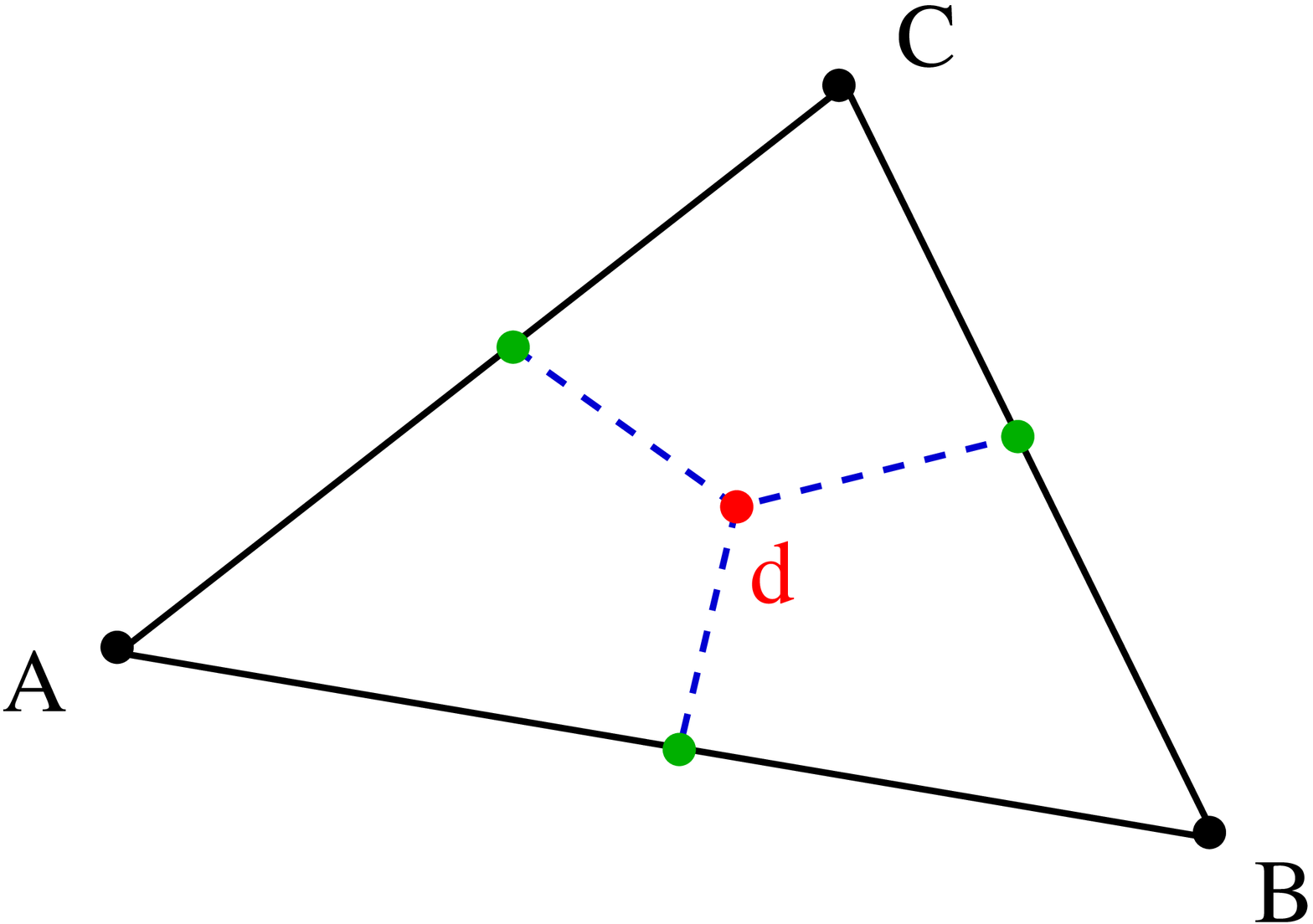}
}
\end{picture}
\caption{\label{f:SquaredTriangle} Each triangle in a triangulation of
  a two-dimensional manifold is refined by the addition of extra
  vertexes and edges to produce three quadrilaterals.  This is done by
  first adding as new vertexes the midpoints of each edge, i.e. the
  points ``ab'', ``bc'' and ``ac'' in the figure on the left.  Next the
  centroid of the triangle, i.e. the point ``d'' in the figure on the
  right, is also added as a new vertex.  Finally the line segments
  that join ``d'' to the midpoints ``ab'', ``bc'', and ``ac'', the
  dashed lines in the figure on the right, are added as new edges.}
\end{figure}

This subsection considers the question of whether two- and
three-dimensional manifolds with arbitrary topologies admit multi-cube
representations.  The first step is to show  that every two-manifold
is homeomorphic to a set of squares (i.e. 2-cubes) glued together
along their edges.  The proof is based on the result of
Rad\'{o}~\cite{Rado1925, Moise1977} that all two-dimensional manifolds
admit triangulations, i.e. that any two-manifold is homeomorphic to a
set of triangles glued together along their edges.  It is easy to show
that a simple refinement of any triangulation on a two-dimensional
manifold produces a multi-cube representation of that manifold.  As
illustrated in Fig.~\ref{f:SquaredTriangle}, let points ``A'', ``B'',
and ``C'' denote the vertexes of one of the triangles in the
triangulation.  Add the midpoints of each edge of this triangle as
additional vertexes, labeled ``ab'', ``bc'', and ``ac'' in
Fig.~\ref{f:SquaredTriangle}.  Next, add the centroid of the triangle,
the point labeled ``d'', and finally add as additional edges the line
segments that connect ``d'' with the midpoints ``ab'', ``bc'' and
``ac''.  The resulting complex consists of three quadrilaterals.  When
all of the triangles in a given triangulation are refined in this way,
the result is a multi-cube representation of the two-manifold.  The
refinement consists of a set of quadrilaterals that are glued together
edge to edge.  Since the additional edge vertexes, ``ab'', etc. are
always added at the geometrical midpoints, the edges of neighboring
quadrilaterals constructed in this way will always coincide.  These
quadrilaterals are homeomorphic to squares (2-cubes).  So the
topological structure of a two-manifold can be thought of as a
collection of non-overlapping 2-cubes that cover the manifold, plus a
set of maps that identify the edges of adjoining 2-cubes.

A similar argument shows that every three-dimensional manifold has a
multi-cube representation, i.e. that every three-dimensional manifold
is homeomorphic to a set of non-overlapping ``distorted'' cubes glued
together at their faces.  The proof is based on a result of
Moise~\cite{Moise1977, Moise1952} that all three-dimensional manifolds
admit triangulations by tetrahedrons, i.e. that any three-dimensional
manifold is homeomorphic to a set of non-overlapping tetrahedrons glued
together at their faces.  It is easy to show that any tetrahedron can
be decomposed into four ``distorted'' cubes glued together at their
faces.  (The term distorted cube is used here to describe a solid
having six faces, each of which is a plane quadrilateral.)  Distorted
cubes are homeomorphic to geometrical cubes.  It follows that every
triangulation of a three-manifold can be refined (by adding
appropriate vertexes, edges and faces) to obtain a multi-cube
representation, i.e. a set of non-overlapping distorted cubes glued
together at their faces.  This argument demonstrates the existence of
multi-cube representations for any three-dimensional manifold.

\begin{figure}[ht]
\begin{picture}(0,230)(0,75)
\put(15,180){
\includegraphics[width=2.in]{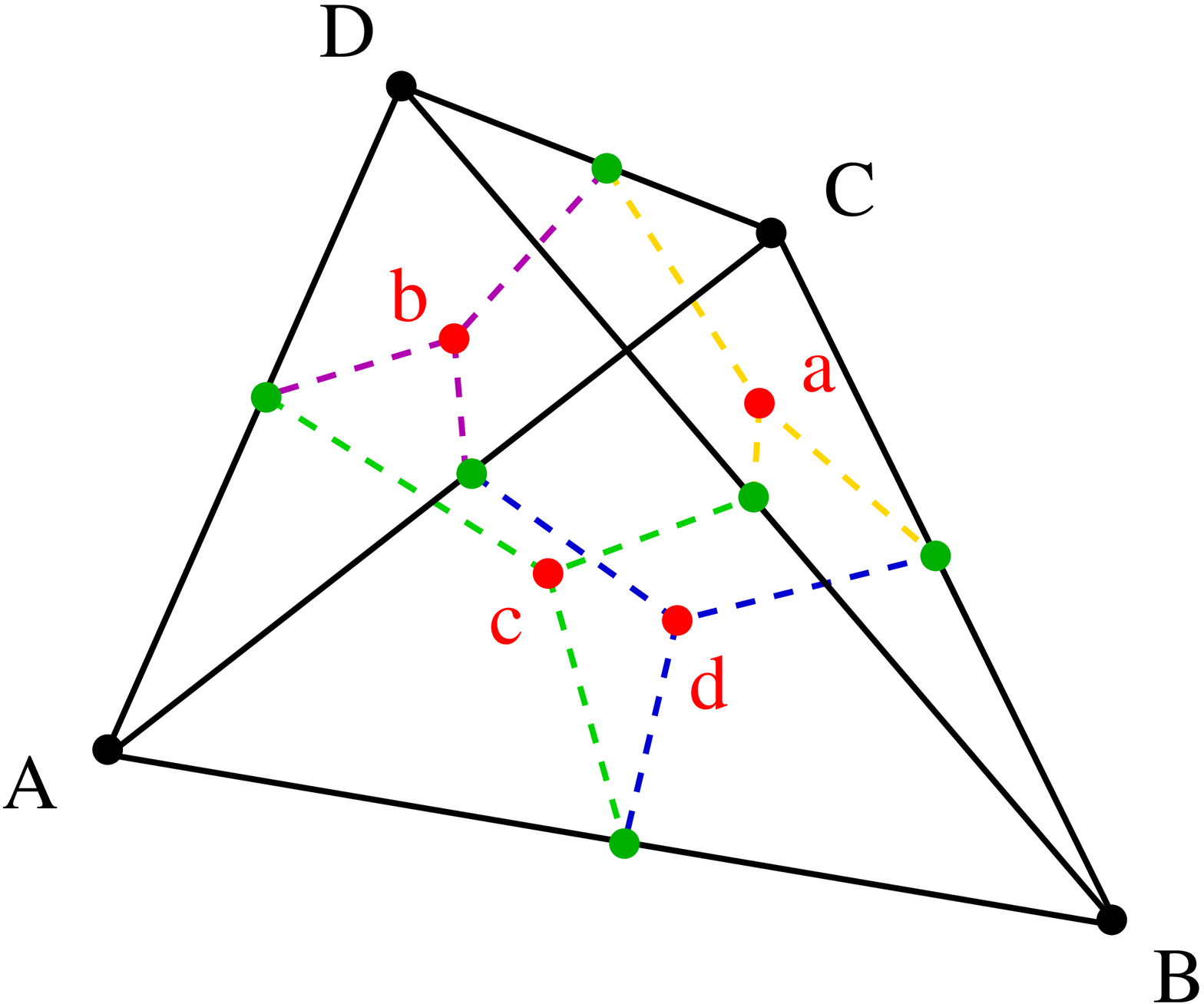}
} 
\put(215,180){
\includegraphics[width=2.in]{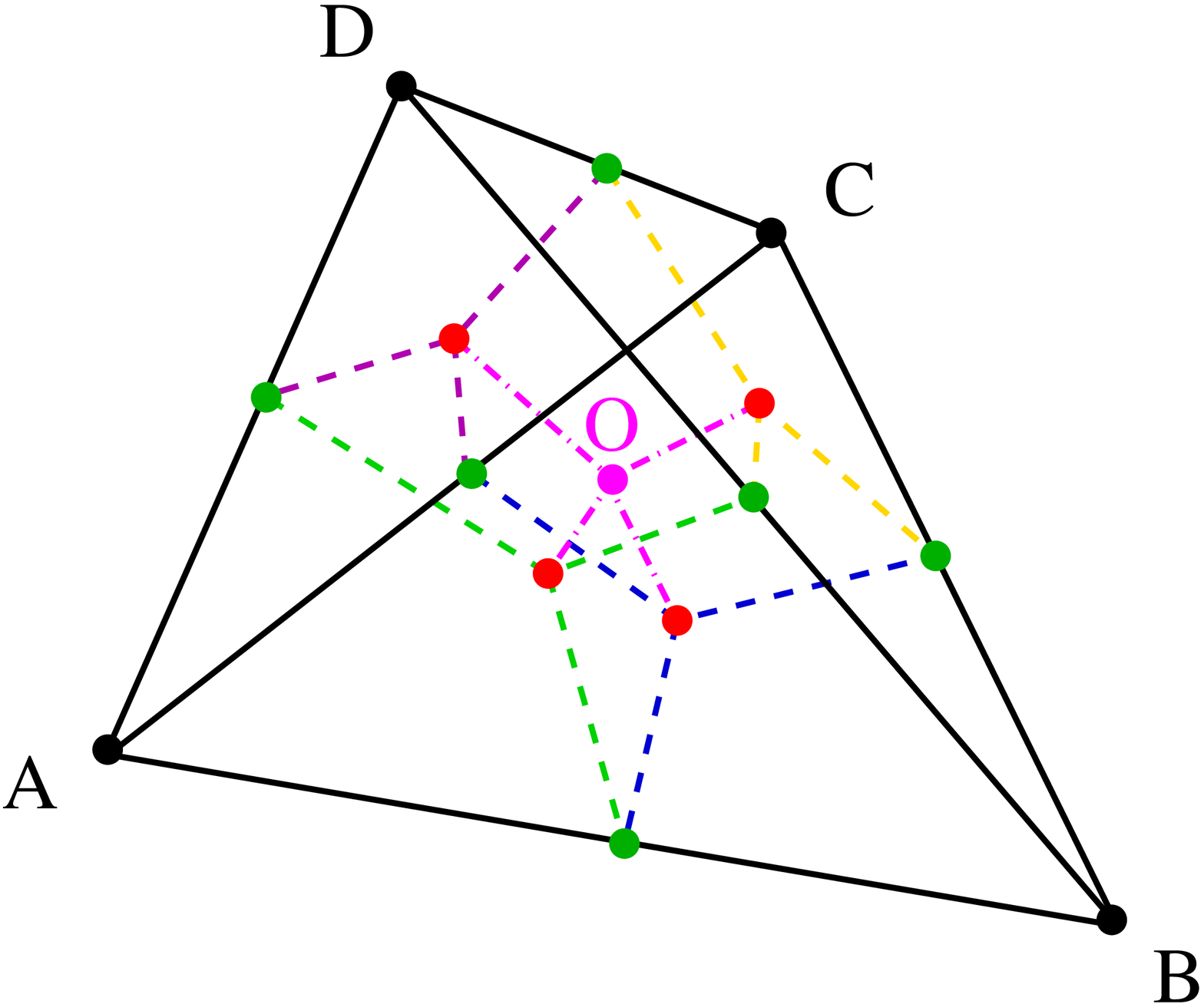}
}
\put(145,80){
\includegraphics[width=2.in]{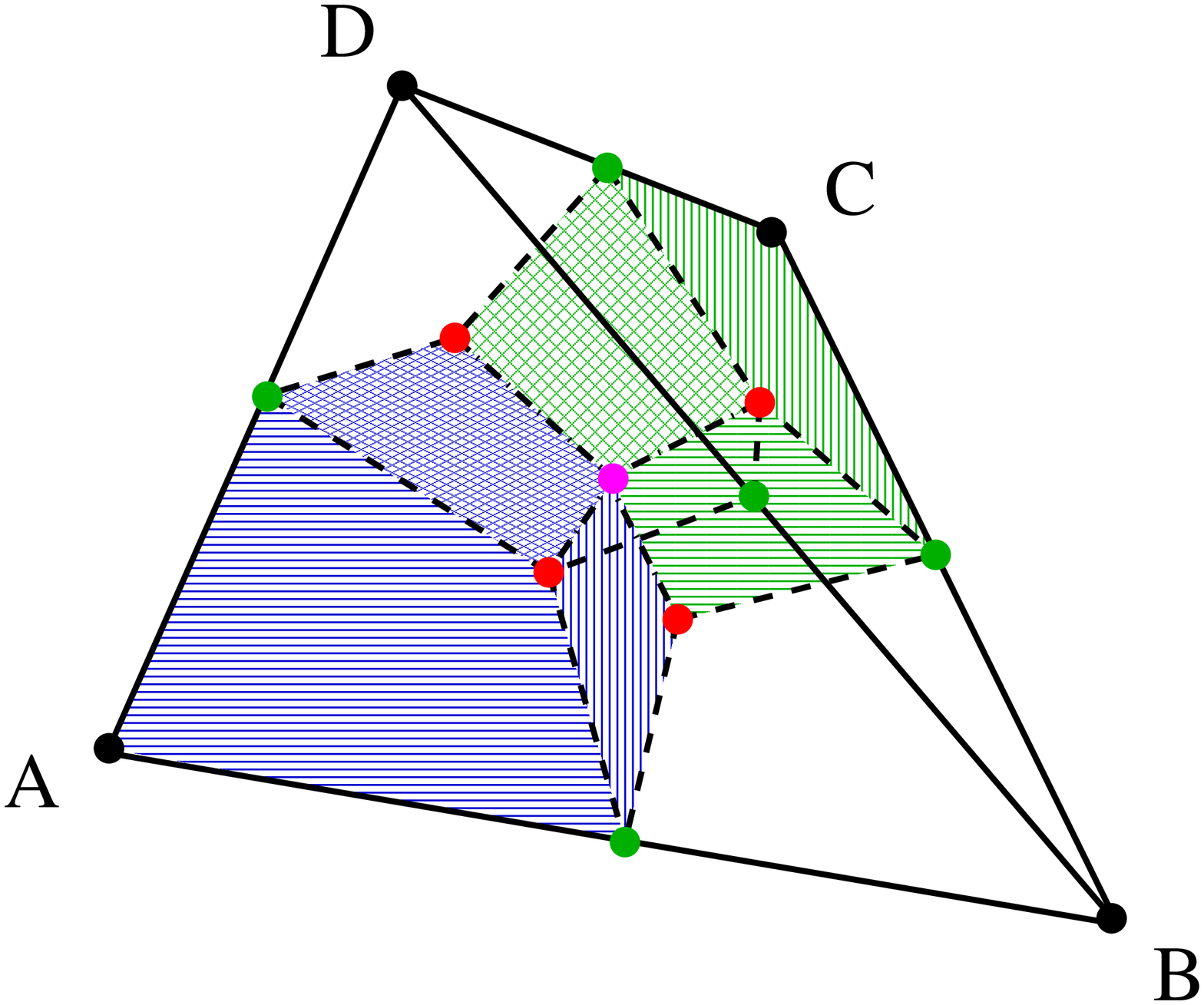}
} 
\end{picture}
\caption{\label{f:CubedTetrahedron} Top Left: Label the vertexes of
  the tetrahedron ``A'', ``B'', ``C'' and ``D''.  Add vertexes at the
  midpoints of each edge, and additional vertexes at the centroid of
  each face of the tetrahedron, labeled ``a'' for the centroid of face
  ``BCD'', ``b'' for face ``ACD'', etc.  Also add additional edges
  (shown as dashed line segments) connecting each centroid to the
  midpoint of each adjoining edge.  Top Right: Add one additional
  vertex, labeled ``O'' at the centroid of the tetrahedron.  Add
  additional edges (shown as dash-dot line segments) that connect
  ``O'' to the centroids of each face, and six additional faces that
  include ``O'' as a vertex.  Bottom: Four ``distorted'' cubes that
  make up the tetrahedron are illustrated.  The two cubes adjacent to
  vertexes ``A'' and ``C'' are shown with opaque shaded faces, while
  the faces of the cubes adjacent to ``B'' and ``D'' are transparent.}
\end{figure}

The key to this argument is the representation of a single tetrahedron
as four distorted cubes glued together.  This can be done by refining
the tetrahedron through the addition of vertexes, edges and faces as
summarized in Fig.~\ref{f:CubedTetrahedron}.  Begin with a tetrahedron
with vertexes labeled ``A'', ``B'', ``C'' and ``D''.  First add
vertexes to the midpoints of each edge, plus vertexes to the centroids
of each face, the points ``a'', ``b'', ``c'' and ``d'' shown in the
top left of Fig.~\ref{f:CubedTetrahedron}.  Adding the extra edges
connecting ``a'', ``b'', ``c'' and ``d'' to the midpoints of each edge
of the original tetrahedron completes the decomposition of each face
into a set of distorted squares.  Add one last vertex at the centroid
of the tetrahedron, labeled ``O'' in the top right of
Fig.~\ref{f:CubedTetrahedron}.  Connect ``O'' to the facial centroids,
``a'', ``b'', ``c'' and ``d'', by adding the edges shown as dash-dot
line segments in the top right of Fig.~\ref{f:CubedTetrahedron}.
Finally add the six internal quadrilateral faces that include the
point ``O'' as an edge vertex.  These additional vertexes, edges, and 
faces divide the tetrahedron into four volume regions (one adjacent to
each tetrahedron vertex).  The bottom of Fig.~\ref{f:CubedTetrahedron}
shows these four regions more clearly.  The regions adjacent to the
vertexes ``A'' and ``C'' are shown with opaque faces, while those
adjacent to ``B'' and ``D'' are shown with transparent faces.

Each of the four volume regions constructed above has six faces, and
each of these faces has four edges and four vertexes.  These faces are
therefore quadrilaterals.  It only remains to show that these
quadrilaterals are planar.  Call two edges of the original tetrahedron
``complimentary'' if they do not intersect at a vertex, e.g. the edges
``AC'' and ``BD'' are complimentary.  Now consider the six bisecting
planes of the tetrahedron, each one formed by an edge and the midpoint
of the complementary edge of the tetrahedron.  Each bisecting plane
passes through the midpoint of the complementary edge, the centroid
``O'', as well as the facial centroids of the two faces adjacent to
the complementary edge.  For example, the bisecting plane formed
by the edge ``AC'' and midpoint ``bd'' intersects ``O'' as well as the
facial centroids ``a'' and ``c''.  The quadrilateral formed by the
vertexes ``bd'', ``a'', ``O'', and ``c'' is therefore a planar
quadrilateral.  It follows that each of the faces of the four volume
regions is a planar quadrilateral, and therefore each volume region is
a distorted cube.
 
The vertexes added in this construction were placed at the geometric
centroids of the triangular faces, and at the centroid of the original
tetrahedron.  The edges added in this construction were also placed in
geometrically determined ways: all of them along one of the bisecting
planes of each edge of the original tetrahedron.  These geometrically
constructed features will therefore match on the triangular boundaries
between neighboring tetrahedrons in any triangulation of a
three-dimensional manifold.  It follows that the distorted cubes
constructed in this way will match face-to-face across all the
tetrahedron boundaries as required for a multi-cube representation of
the manifold.

\subsection{Infrastructure for Multi-Cube Manifolds}
\label{s:InfrastructureForMultiCubes}

Now turn to the problem of finding a systematic way of constructing
multi-cube manifolds.  The goal is to develop methods that can be used
as part of the computational infrastructure for solving systems of
partial differential equations on such manifolds.  The discussion here
is focused on three-dimensional manifolds $\Sigma$, but
generalizations to other dimensions should be fairly straightforward.
Let ${\cal B}_A$ denote a collection of geometrical cubic regions in
$R^3$.  The subscript ${\scriptstyle A} = \{1,...,N\}$ is used to
label the individual regions.\footnote{The term region in this paper
  is used to refer to the cubes ${\cal B}_A$ that form the
  basic topological structure of the manifold.  It might be useful for
  computational efficiency to subdivide some (or all) of the cubic
  regions into a collection of smaller cubes, e.g. by cutting a cubic
  region into two, four, or eight smaller cubes.  Those smaller cubic
  subsets of the ${\cal B}_A$ are referred to as subregions.}  These
cubes are used here as the domains of coordinate charts for the
multi-cube representation of $\Sigma$.  Let $\Psi_A$ denote the
invertible coordinate map that takes the region ${\cal B}_A$ into a
subset of $\Sigma$: $\Psi_A({\cal B}_A)\subset \Sigma$.  It will be
useful to denote the boundary faces of these regions in $R^3$ as
$\partial_\alpha{\cal B}_A$, where $\alpha=\pm x$ denotes the faces
intersecting the $\pm x$ axes, $\alpha=\pm y$ the faces intersecting
the $\pm y$ axes, etc.

The discussion above shows that every three-manifold can be covered by
a collection of non-overlapping cubes: $\cup_{A}\Psi_A({\cal
  B}_A)=\Sigma$.  Non-overlapping here means that the images of the
regions are non-intersecting, $\Psi_A({\cal B}_A)\cap\Psi_B({\cal
  B}_B) =\emptyset$, for points in the interiors of ${\cal B}_A$ and
${\cal B}_B$ when $\scriptstyle{A}\neq\scriptstyle{B}$.  It is
convenient to choose the regions ${\cal B}_A$ in $R^3$ to be scaled so
they all have the same size $L$, and are all oriented along the same
global Cartesian coordinate axes in $R^3$.  In this case the region
${\cal B}_A$ is completely determined therefore simply by specifying
the location of its center $\vec c_A=(c{}^x{}_A,c{}^y{}_A,c{}^z{}_A)$
in $R^3$.  It is also convenient to arrange the regions ${\cal B}_A$
in $R^3$ so they intersect (if at all) in $R^3$ only at points on
faces whose images also intersect in $\Sigma$.  In the multi-cube
representations of manifolds satisfying these conditions, each point
in the interior of the regions represents a unique point in $\Sigma$,
and each point in $\Sigma$ is the image of at least one point in the
closure of $\cup_A{\cal B}_A$.  The Cartesian coordinates of $R^3$
therefore provide a global way of identifying points in $\Sigma$.
Tensor fields are represented on these multi-cube manifolds by giving
the values of their components (expressed in the coordinate basis of
$R^3$) as functions of these global Cartesian coordinates.

A multi-cube manifold consists of a set of cubic regions, ${\cal B}_A$
for ${\scriptstyle A}=\{1, ..., N\}$ that can be specified simply by
giving the locations of their centers $\vec c_A$, along with a set of
rules that determine how the faces of these cubes are to be identified
with one another.  When points on the images of two boundary faces
$\Psi_A(\partial_\alpha{\cal B}_A)$ and $\Psi_B(\partial_\beta{\cal
  B}_B)$ intersect in $\Sigma$, then the associated coordinate charts
provide an invertible map from one boundary face to the other:
$\partial_\alpha{\cal B}_A=\Psi_{B\beta}^{A\alpha}(\partial_\beta{\cal
  B}_B)$ where $\Psi_{B\beta}^{A\alpha}\equiv\Psi^{-1}_A\circ \Psi_B$
for points on the $\partial_\alpha{\cal B}_A$ and $\partial_\beta{\cal B}_B$
faces.  Since the cubes ${\cal B}_A$ have uniform size and orientation
in $R^3$, there are only a small number of simple maps
$\Psi_{B\beta}^{A\alpha}$ needed to represent all the topologically
distinct ways of mapping one face onto another.  It is sufficient to
consider maps that identify the faces of two cubic region first by
rigidly translating so the centers of the faces $\partial_\alpha{\cal
  B}_A$ and $\partial_\beta{\cal B}_B$ coincide, and then rigidly
rotating and/or reflecting to align the two faces in the desired way.
Thus it is sufficient to consider the simple maps
$\Psi_{B\beta}^{A\alpha}$ that take the Cartesian coordinates $x^i_B$
of points in $\partial_\beta{\cal B}_B$ to the Cartesian coordinates
$x^i_A$ of the corresponding points in $\partial_\alpha{\cal B}_A$ in
the following way,
\begin{eqnarray}
x^i_A = c^i_A +f^i_{\alpha} + C_{B\beta\,j}^{A\alpha\, i}(x^j_B
 - c^j_B-f^j_{\beta}).
\label{e:CoordinateMap}
\end{eqnarray}
The vector $\vec c_A+\vec f_{\alpha}$ is the location of the center of
the $\partial_\alpha{\cal B}_A$ face, and ${\mathbf
  C}_{B\beta}^{A\alpha}$ is the combined rotation and reflection
matrix needed to achieve the desired orientation.  Examples of the use
of these methods is given in \ref{s:ExampleRepresentations} where
explicit multi-cube representations are constructed for manifolds with
the topologies $T^3$, $S^2\times S^1$ and $S^3$.

Multi-cube manifolds are specified by giving the list of cubic regions
${\cal B}_A$ needed to cover the manifold, the vectors $\vec c_A$ that
determine the locations of their centers in $R^3$, and the maps
$\Psi_{B\beta}^{A\alpha}$ that determine how the regions are glued
together.  These maps, defined in Eq.~(\ref{e:CoordinateMap}), depend
on the vectors $\vec c_A$ and $\vec f_\alpha$, and the matrix ${\mathbf
  C}_{B\beta}^{A\alpha}$, so these quantities must all be specified to
determine each map.  The vector $\vec f_\alpha$ is the position of the
center of the $\alpha$ face relative to the center of the
region. Since the cubic regions are chosen to have uniform sizes and
orientations, $\vec f_\alpha$ has the same form in each cubic region:
\begin{eqnarray}
\vec f_{\pm x}&=&{\scriptstyle \frac{1}{2}}L (\pm 1,0,0),\nonumber\\
\vec f_{\pm y}&=&{\scriptstyle \frac{1}{2}}L (0,\pm 1,0),\\
\vec f_{\pm z}&=&{\scriptstyle \frac{1}{2}}L (0,0,\pm 1),\nonumber
\end{eqnarray}
where $L$ is the size of the cubes.  Since all of the cubic regions
are aligned, the class of possible rotations and reflections needed
for ${\mathbf C}_{B\beta}^{A\alpha}$ is quite small.  These can all be
constructed by combining 90-degree rotations about the normal to one
of the faces, ${\mathbf R}_\alpha$, with mirror reflections about some
(possibly different) direction, ${\mathbf M}_\beta$.
Table~\ref{t:TableI} gives explicit expressions for the matrices that
describe these elementary rotations and reflections in three
dimensions.
\begin{table}[ht]
\renewcommand{\arraystretch}{1.2}
\begin{center} 
\caption{Elementary Transformations
\label{t:TableI}}
\begin{tabular}{c|cccc}   
\hline\hline
&&$\alpha=\pm x$ &$\alpha=\pm y$ &$\alpha=\pm z$\\ 
\hline
&&&&\\
${\mathbf R}_\alpha$ 
&& $\left(
\begin{array}{ccc}
1 & 0 & 0\\
0 & 0 & \mp 1\\
0 & \pm 1 & 0
\end{array}
\right)$

& $\left(
\begin{array}{ccc}
0 & 0 & \pm 1\\
0 & 1 & 0\\
\mp 1 & 0 & 0
\end{array}
\right)$ 

&$\left(
\begin{array}{ccc}
0 & \mp 1 & 0\\
\pm 1 & 0 & 0\\
0 & 0 & 1
\end{array}
\right)$
\\ 
&&&&\\
${\mathbf M}_\alpha$ 
&&
$\left(
\begin{array}{ccc}
-1 & 0 & 0\\
0& 1 & 0\\
0 & 0 & 1

\end{array}
\right)$
&
$\left(
\begin{array}{ccc}
1 & 0 & 0\\
0& -1 & 0\\
0 & 0 & 1
\end{array}
\right)$
&
$\left(
\begin{array}{ccc}
1 & 0 & 0\\
0& 1 & 0\\
0 & 0 & -1
\end{array}
\right)$\\
&&&\\
\hline\hline
 \end{tabular}
\end{center}
\end{table}
The most general transformation of one face onto another can be
constructed by taking products of these elementary transformations.
The group of possible ${\mathbf C}_{B\beta}^{A\alpha}$ in three
dimensions generated in this way is therefore the octahedral symmetry
group, $O_h$, which has 48 distinct elements~\cite{Hamermesh1962}.
The orientation preserving subgroup generated by the rotations alone
has 24 elements.  Note that ${\mathbf R}_{\alpha}\cdot {\mathbf
  R}_{-\alpha}={\mathbf R}_\alpha^4 = {\mathbf M}_\alpha^2 = {\mathbf
  I}$, where ${\mathbf I}$ is the identity matrix.  Since the number
of possible maps $\Psi_{B\beta}^{A\alpha}$ constructed in this way is
so small, it is easy to write a flexible code that is capable of
setting up the multi-cube structures and all the needed gluing maps
for three-manifolds with arbitrary topologies.

\section{Specifying Differential Structures on Multi-Cube Manifolds}
\label{s:DifferentialStructure}

This section describes a practical and efficient way to define $C^k$
differential structures on multi-cube manifolds.  It is useful to
begin with a brief discussion of the traditional way such structures
are defined.  The differential structure on a manifold provides the
framework needed to represent differentiable scalar and tensor fields
on that manifold.  The usual method of specifying a differential
structure is to cover the manifold with a set of overlapping domains
${\cal D}_A$, and set of maps $\Upsilon_A$ that assign coordinates to
the points in each domain: $\Upsilon_A^{-1}({\cal D}_A)\subset R^n$.
These coordinate maps provide a differential structure for the
manifold if they have the property that the composition maps
$\Upsilon_B^A = \Upsilon_A^{-1} \circ \Upsilon_B$ are differentiable
(or $C^{k+1}$) transformations from the coordinates of one patch to
the other for points in the overlap ${\cal D}_A\cap {\cal D}_B$.
The Jacobian matrices associated with these coordinate transformations
$J^{Ai}_{Bj}=\partial x_A^i/\partial x_B^j$ determine the
transformations for $C^k$ differentiable tensors from one coordinate
representation to another in these overlaps.

It is possible to use the traditional method of defining differential
structures on multi-cube manifolds, but to do so requires that
non-trivial additional structures must be added to the basic
multi-cube construction (since the domains that define that basic
structure do not overlap).  The most straightforward approach would be
to require that each multi-cube manifold be provided with an
additional set of overlapping domains ${\cal D}_A\supset \Psi_A({\cal
  B}_A)$ and a set of $C^{k+1}$ related coordinate maps $\Upsilon_A$
for the new overlapping ${\cal D}_A$ domains.  An alternative, more
minimalist, approach would be to require that suitable Jacobian
matrices $J^{A\alpha i}_{B\beta j}$, in addition to the connection
maps $\Psi_{B\beta}^{A\alpha}$, be provided on each interface between
regions in multi-cube manifolds.  This minimal structure would provide
the transformations needed to define differentiable scalar and
continuous tensor fields on these manifolds.  If $C^{k+1}$
differentiable scalars or $C^k$ differentiable tensor fields are
needed, then in addition to $J^{A\alpha i}_{B\beta j}$, all of their
$k^\mathrm{th}$ order derivatives $\partial^k_BJ^{Ai}_{Bj}$ would also
have to be specified on each interface between regions.

\begin{figure}[ht]
\centerline{\includegraphics[width=2.0in]{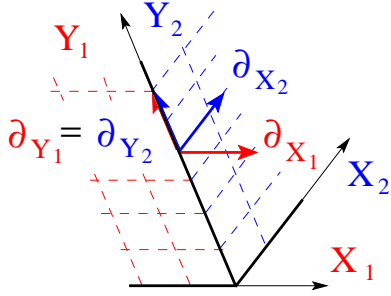}}
\caption{\label{f:MeshBoundary} Maps $\Psi_A$ define continuous but
  (typically) non-differentiable transitions between cubic
  regions.  This example shows that the basis vectors tangent to the
  boundary, $\partial_{Y_1}$ and $\partial_{Y_2}$, are continuous,
  while those not tangent to the boundary, $\partial_{X_1}$ and
  $\partial_{X_2}$, are not.  }
\end{figure}

It might seem redundant and unnecessary to require that the Jacobian
matrices $J^{A\alpha i}_{B\beta j}$ and their derivatives be specified
on the interfaces in multi-cube manifolds, in addition to the
interface coordinate maps $\Psi^{A\alpha}_{B\beta}$ defined in
Eq.~(\ref{e:CoordinateMap}).  After all, the Jacobian matrices
associated with those interface maps, $J^{A\alpha i}_{B\beta j}=
C^{A\alpha i}_{B\beta j}$, and their derivatives,
$\partial_{Bk}J^{A\alpha i}_{B\beta j}= \partial_{Bk}C^{A\alpha
  i}_{B\beta j}=0$, could be used to transform tensor fields at the
boundary interfaces.  Unfortunately it is easy to see that the
coordinate maps $\Psi_A$ used in Sec.~\ref{s:TopologicalStructure} to
construct the multi-cubes are not suitable for constructing a global
$C^k$ differential structure on most manifolds.  If they were, the
basis vectors $\partial_{Ai}$ associated with these coordinates would
be smooth global non-vanishing vector fields.  These vector fields
could be used in this case to construct a global smooth flat metric on
the manifold.  Since most manifolds do not admit global flat metrics,
the existence of a complete set of smooth non-vanishing coordinate
vector fields can not exist on most manifolds.
Figure~\ref{f:MeshBoundary}, drawn from the perspective of a smooth
coordinate patch that covers both sides of an interface boundary,
illustrates how the multi-cube coordinates in neighboring regions can
be continuous while failing to be differentiable across region
boundaries.  The coordinate region ${\cal B}_1$ on the left, matches
to coordinate region ${\cal B}_2$ on the right across the $X_1=X_2$
interface in Fig.~\ref{f:MeshBoundary}.  The coordinate vectors
tangent to this interface, e.g. $\partial_{Y_1}$ and $\partial_{Y_2}$,
are continuous across this interface, while those not tangent to the
boundary, i.e.  $\partial_{X_1}$ and $\partial_{X_2}$, are
discontinuous there.

Both approaches described above for specifying differential structures
on a multi-cube manifolds require that a great deal of extra structure
be provided.  This paper proposes a third, more elegant and more
efficient, approach that can be incorporated more easily into the
computational infrastructure for solving partial differential
equations numerically.  Every manifold with a $C^{k+1}$ differential
structure admits a symmetric positive definite $C^k$ differentiable
metric tensor $g_{ij}$.  The method proposed here for specifying the
global differential structure on a multi-cube manifold requires that
the components of (any) one of these $C^k$ differentiable reference
metrics, $g_{ij}$, be provided in the global Cartesian coordinate
basis used to define the multi-cube manifold.  The components of this
reference metric $g_{ij}$ will be $C^k$ functions of the multi-cube
Cartesian coordinates within each region ${\cal B}_A$, but will (in
general) be discontinuous across the interfaces between regions.  The
only requirement on this reference metric is that it must be
sufficiently smooth, $C^k$, when represented in a global $C^{k+1}$
coordinate atlas.  The $C^{k+1}$ coordinate charts $\Upsilon_A$
themselves need not be given as part of the specification of the
multi-cube manifold.  Their only use in this method is to ensure {\it
  a priori} that the reference metric meets the needed smoothness
requirements.

Once a suitable reference metric $g_{ij}$ is provided, it is
straightforward to construct the Jacobian matrices $J^{A\alpha
  i}_{B\beta j}$ and the dual Jacobian matrices $J_{A\alpha
  i}^{*B\beta j}$ needed to transform continuous tensor fields across
the interface boundaries in multi-cube manifolds.  Assume that the
$\partial_\alpha{\cal B}_A$ boundary of region ${\cal B}_A$ is
identified with the $\partial_\beta{\cal B}_B$ boundary of region
${\cal B}_B$ by the map $\Psi^{A\alpha}_{B\beta}$ given in
Eq.~(\ref{e:CoordinateMap}).  The transformation taking the region
${\cal B}_B$ representation of a vector $v^i_B$ into the region ${\cal
  B}_A$ representation $v_A^i$ at one of these identified boundary
points is an expression of the form
\begin{eqnarray}
v_A^i = J^{A\alpha i}_{B\beta j}v_B^j,
\label{e:Jdef}
\end{eqnarray}
where $J^{A\alpha i}_{B\beta j}$ is in effect the Jacobian matrix of
the transformation.  The analogous transformation law for covectors
$w_{Bi}$ is,
\begin{eqnarray}
w_{Ai} &=& J_{A\alpha i}^{*B\beta j}\, w_{Bj} ,
\label{e:dualJdef}
\end{eqnarray}
where $J_{A\alpha i}^{*B\beta j}$ is in effect the dual Jacobian
matrix.

Let $g_{Aij}$ denote the coordinate components of the reference
metric in the multi-cube coordinate basis of region ${\cal B}_A$, and
let $n_{A\alpha i}$ denote the outward directed normal covector to the
surface $\partial_\alpha{\cal B}_A$.  This interface is a surface of
constant coordinate $x_A^\alpha$, so the geometrical normal covector is
proportional to $\partial_{Ai}x_A^\alpha$.  The normal covector
is therefore given by
\begin{eqnarray}
n_{A\alpha i}=\frac{\pm \partial_{Ai}x^\alpha_A}
{\sqrt{g_A^{jk}\partial_{Aj}x^\alpha_A\partial_{Ak}x^\alpha_A}},
\end{eqnarray}
where $g_A^{ij}$ is the inverse of the reference metric $g_{Aij}$.
The sign is chosen in this expression to make $n_{A\alpha i}$ the
outgoing unit normal. The unit normal vector $n_{A\alpha}^i$ is
related to $n_{A\alpha i}$ by $n_{A\alpha}^i=g_A^{ij}n_{A\alpha j.}$.

The Jacobian matrices needed to transform vectors and covectors (and
therefore any type of tensor field) across boundary interfaces are
simple functions of the quantities $C^{A\alpha i}_{B\beta j}$ and
$C_{A\alpha i}^{B\beta j}$ (which define the identification maps
$\Psi^{A\alpha}_{B\beta}$), as well as the normals to the boundary
surface, $n_{A\alpha}^i$, $n_{A\alpha i}$, $n_{B\beta}^i$ and
$n_{B\beta i}$:
\begin{eqnarray}
J^{A\alpha i}_{B\beta j} &=& C^{A\alpha i}_{B\beta k}
\left(\delta^k_j - n^k_{B\beta} n_{B\beta j}\right) - n^i_{A\alpha} n_{B\beta j},
\label{e:Tdef}\\
J_{A\alpha i}^{* B\beta j} &=& 
\left(\delta_i^k - n_{A\alpha i} n_{A\alpha}^k\right)C^{B\beta j}_{A\alpha k}
 - n_{A\alpha i} n_{B\beta}^j.
\label{e:dualTdef}
\end{eqnarray}
The Jacobian matrices defined in Eqs.~(\ref{e:Tdef}) and
(\ref{e:dualTdef}) are the unique ones with the properties:\hfill\break 
{\it a)} They map the geometrical normals $n_{B\beta}^j$ into $-n_{A\alpha}^i$
and $n_{B\beta j}$ into $-n_{A\alpha i}$,
\begin{eqnarray}
n^i_{A\alpha} &=& - J^{A\alpha i}_{B\beta j}\,n_{B\beta}^j,\\
n_{A\alpha i} &=& - J_{A\alpha i}^{* B\beta j} n_{B\beta j},
\end{eqnarray}
(i.e. the outward directed normal of one region is identified with the
inward directed normal of its neighbor). \hfill\break
{\it b)} The Jacobian matrix
$J^{A\alpha i}_{B\beta j}$ transforms any vector $t^i$ tangent to the
boundary (i.e. any vector satisfying $t^i n_i=0$) using the continuity
of the $\Psi^{A\alpha}_{B\beta}$ maps:
\begin{eqnarray}
t^{\,i}_A &=&J^{A\alpha i}_{B\beta j}\,t^{\,j}_B\,\,\, =\,\,\,
C^{A\alpha i}_{B\beta j}\,t^{\,j}_B.
\end{eqnarray}
{\it c)} The Jacobian matrix $J^{A\alpha i}_{B\beta j}$ and its dual
$J_{A\alpha i}^{* B\beta j}$ are inverses
\begin{eqnarray}
\delta^{Ai}_{Aj} = J^{A\alpha i}_{B\beta k}\,
J_{A\alpha j}^{* B\beta k}.
\end{eqnarray}
This last property ensures that tensor contractions and traces
transform properly under these boundary interface mappings.

The Jacobian matrices constructed in Eqs.~(\ref{e:Tdef}) and
(\ref{e:dualTdef}) using the identification maps
$\Psi^{A\alpha}_{B\beta}$ and the reference metric $g_{ij}$ define the
transformations needed to connect arbitrary tensor fields across the
interface boundaries of multi-cube manifolds.  These transformations
make it possible therefore to define what it means for a global tensor
field to be continuous on multi-cube manifolds: A tensor field is
continuous on a multi-cube manifold if its multi-cube coordinate
components are continuous within each region ${\cal B}_A$, and if its
multi-cube coordinate components at each interface boundary point are
equal to the transform of its components from the neighboring region.

The reference metric can also be used to define a smooth connection
\begin{eqnarray}
\Gamma^i_{jk} = {\scriptstyle \frac{1}{2}}g^{i\ell}\left(\partial_jg_{\ell k}
+\partial_kg_{\ell j}-\partial_\ell g_{j k}\right),
\label{e:GammaDef}
\end{eqnarray}
that can be used to define a covariant derivative operator $\nabla_i$.
This covariant derivative is related to the coordinate partial
derivatives (within each region ${\cal B}_A$) by the usual expressions
for the case of vectors and covectors:
\begin{eqnarray}
\nabla_i v^j &=& \partial_i v^j + \Gamma^j_{ik} v^k,\\
\nabla_i w_j &=& \partial_i w_j - \Gamma^k_{ij}w_k.
\end{eqnarray}
The covariant gradients of tensors, e.g. $\nabla_i v^j$ and $\nabla_i
w_j$, are themselves tensor fields.  Therefore they are transformed at
interface boundaries using the Jacobian matrices defined in
Eqs.~(\ref{e:Tdef}) and (\ref{e:dualTdef}) as well.  Thus, for
example, the gradients of vectors and covectors transform as,
\begin{eqnarray}
\nabla_{Ai} v^j_A &=& J_{A\alpha i}^{*B\beta k}J^{A\alpha j}_{B\beta \ell}
\nabla_{Bk} v^\ell_B,
\label{e:dTensorTransformVec}
\\
\nabla_{Ai} w_{Aj} &=& J_{A\alpha i}^{*B\beta k}J_{A\alpha j}^{*B\beta \ell}
\nabla_{Bk} w_{B\ell}.
\label{e:dTensorTransformCoVec}
\end{eqnarray}
Using these transformation laws it is straightforward to define what it
means for a global tensor field to be differentiable on a multi-cube
manifold: A tensor field is differentiable if the tensor and its
covariant gradient are continuous everywhere including across all
multi-cube interfaces.  The concept of $C^k$ tensors can be built up
in a straightforward way simply by taking $k^\mathrm{th}$ order
covariant gradients of tensors and demanding that the tensor and all
gradients up through $k^\mathrm{th}$ order be continuous global tensor
fields.

The addition of a smooth (i.e. $C^k$ differentiable) positive definite
reference metric $g_{ij}$ therefore provides all the additional
information needed to define a global $C^k$ differential structure on
any multi-cube manifold.
  

\section{Interface Boundary Conditions for Multi-Cube Manifolds}
\label{s:BoundaryConditions}

The multi-cube representations of manifolds provide a practical
framework in which to solve systems of partial differential equations
numerically on manifolds with non-trivial spatial topologies.  The
idea is to solve those equations on each of the cubic regions ${\cal
  B}_A$ separately, using boundary conditions on the faces
$\partial_\alpha{\cal B}_A$ that ensure the combination of local
solutions from each region satisfies the system of equations
globally---including at the boundaries.  Solving differential
equations using multi-patch methods is a common practice in
computational physics on manifolds that are subsets of
$R^3$~\cite{Thornburg2000,Thornburg2004a,Lehner:2005bz, 
Schnetter:2006pg, Pazos:2009vb,Pollney:2009yz,Korobkin2011}.  Such
methods are used for example in the pseudo-spectral code SpEC
(developed by the Caltech/Cornell numerical relativity
collaboration~\cite{Kidder2000a, Pfeiffer2003,Scheel2006, Scheel2009,
  Szilagyi:2009qz}) to solve Einstein's equations.  The multi-cube
framework developed here extends the class of problems accessible to
such codes by allowing them to solve problems on computational domains
that can not be covered by a single global coordinate chart.  This
generalization provides a method of solving differential equations on
two-dimensional and three-dimensional manifolds with arbitrary
topologies, in addition to a very large class of higher dimensional
manifolds.  The code changes needed to implement these more general
multi-cube methods require fairly minor generalizations of the way
boundary conditions are imposed at the interfaces between cubic
regions in standard multi-patch codes.  The needed generalizations are
described here in some detail for second-order quasi-linear
strongly-elliptic and first-order symmetric-hyperbolic systems of
equations.

\subsection{Interface Boundary Conditions for Elliptic Systems}
\label{s:EllipticBC}

A second-order quasi-linear strongly-elliptic system of equations for
a collection of tensor fields $u^{\cal A}$ can be written in the form
\begin{eqnarray}
\nabla_j\left[ M^{jk \cal A}{}_{\cal B}{}(\mathbf{u}) \nabla_k u^{\cal B}\right]
= F^{\cal B}(\mathbf{u},\mathbf{\nabla u}),
\label{e:EllipticSystem}
\end{eqnarray}
where $\nabla_i$ is some covariant derivative operator, $M^{jk\cal
  A}{}_{\cal B}(\mathbf{u})$ may depend on the fields but not their
derivatives, and $F^{\cal B}(\mathbf{u},\mathbf{\nabla u})$ may depend
on the fields and their first derivatives.  The script indexes
${\scriptstyle {\cal A}}$, ${\scriptstyle {\cal B}}$, ${\scriptstyle
  {\cal C}}$, ... in these expressions label the components of the
collection of tensor fields that make up $u^{\cal A}$.  Such a system
is strongly elliptic if there is a positive definite metric on the
space of fields, $S_{\cal AB}$, a positive definite spatial metric,
$g^{ij}$, on the manifold (e.g. the reference metric used to define
the multi-cube structure) and a positive constant, $C>0$, such that
\begin{eqnarray}
w_j w_kM^{jk\cal C}{}_{\cal A} S_{\cal CB}\,v^{\cal A} v^{\cal B} 
\geq C \,g^{jk} w_j w_k\,S_{\cal AB}\, v^{\cal A} v^{\cal B} 
\end{eqnarray}
for every  $v^{\cal A}$ and every $w_j$~\cite{McLean2000}.

All differentiable soltuions to second-order elliptic systems of this type 
are smooth, assuming the quantities $M^{jk\cal
  A}{}_{\cal B}$ and $F^{\cal B}$ are smooth~\cite{McLean2000}.
Boundary conditions for these equations at internal inter-region
boundaries are therefore quite simple: the solutions $u^{\cal A}$ and
their normal derivatives $n^i\nabla_i u^{\cal A}$ (where $n^i$ is
the normal to the boundary) must be continuous when transformed
appropriately across inter-region boundaries.

These continuity conditions can only be imposed at the interface
boundaries by transforming the fields $u^{\cal A}$ computed in one
region, ${\cal B}_B$, into the tensor basis used by its neighboring
region, ${\cal B}_A$.  The fields $u^{\cal A}$ are (by assumption) a
collection of tensor fields whose components are transformed across
region boundaries using the Jacobian as defined in Eqs.~(\ref{e:Jdef})
and (\ref{e:dualJdef}).  Thus the fields $u^{\cal A}_B$ (expressed in
the tensor basis associated with the coordinates ${x}^i_B$ from the
region ${\cal B}_B$) are related to the fields $u^{\cal A}_A$ (in the
tensor basis associated with the coordinates ${x}^i_A$ from the region
${\cal B}_A$) by a transformation of the form,
\begin{eqnarray}
u^{\cal A}_A
={\cal J}^{\,{\cal A}}{}_{\!\!{\cal B}}\,u^{\cal B}_B,
\label{e:FieldTransformation}
\end{eqnarray}
where ${\cal J}^{\,{\cal A}}{}_{\!\!{\cal B}}$ is the multi-component Jacobian
appropriate for each tensor part of $u^{\cal B}$.  For example, a
system whose fields consist of a scalar, a vector, and a covector
$u^{\cal B}=\{\psi,v^i,w_i\}$, would transform as follows,
\begin{eqnarray}
&&\!\!\!\!\!
{\cal J}^{\,{\cal A}}{}_{\!\!{\cal B}}\,u^{\cal B}_B
=\left\{\psi_B,\,\,
J^{A\alpha j}_{B\beta i}v^i_B,\,\,
J_{A\alpha j}^{*B\beta i}w_{Bi}
\right\}.
\label{e:TensorFieldTransExample}
\end{eqnarray}

The boundary conditions for second-order elliptic systems also place
conditions on the normal derivatives of the fields, $n^i\nabla_{i}
u^{\cal A}$.  The covariant gradient of a tensor field is itself a
tensor field, so these gradients are transformed across region
boundaries by an equation analogous to
Eq.~(\ref{e:FieldTransformation}):
\begin{eqnarray}
\nabla_{Ai}u^{\cal A}_A = J^{*B\beta j}_{A\alpha i}{\cal J}^{\cal A}_{\cal B}
\nabla_{Bj} u^{\cal B}_B.
\label{e:DerivativeFieldTransformation}
\end{eqnarray}
It may be more convenient in some cases to impose the needed continuity
conditions on the partial derivatives, $n^i\partial_i u^{\cal A}$,
rather than the covariant derivatives of the fields, $n^i\nabla_i
u^{\cal A}$.  The interface boundary transformations needed in this
case are easy to obtain from
Eq.~(\ref{e:DerivativeFieldTransformation}): the covariant derivatives
$\nabla_{Ak}$ and $\nabla_{Bk}$ that appear in this condition are
re-expressed in terms of the partial derivatives $\partial_{Ai}$ and
$\partial_{Ai}$, and the connection coefficients $\Gamma^{\,i}_{Ajk}$
and $\Gamma^{\,i}_{Bjk}$.  For the case of vector and co-vector
fields, the resulting partial derivative transformation laws are given
by,
\begin{eqnarray}
\partial_{Ak} v_A^i&=&J^{*B\beta \ell}_{A\alpha k}J^{A\alpha i}_{B\beta j}
\,\partial_{B\ell}v_B^j
+\left( J^{*B\beta \ell}_{A\alpha k}J^{A\alpha i}_{B\beta n}
\,\Gamma^{\,n}_{B\ell j} 
- J^{A\alpha n}_{B\beta j}\,\Gamma^{\,i}_{Akn}\right)v_B^j,\quad
\label{e:partialTensorTransformVec}
\\
\partial_{Ak} w_{Ai}&=&J^{*B\beta \ell}_{A\alpha k}J_{A\alpha i}^{*B\beta j}
\,\partial_{B\ell}v_{Bj}
-\left( J^{*B\beta \ell}_{A\alpha k}J_{A\alpha i}^{*B\beta n}
\,\Gamma^{\,j}_{B\ell n} 
- J_{A\alpha n}^{*B\beta j}\,\Gamma^{\,n}_{Aki}\right)w_{Bj}.\quad
\label{e:partialTensorTransformCoVec}
\end{eqnarray}

The needed interface boundary conditions for second-order elliptic
systems can now be stated precisely: Let ${\cal B}_A$ and ${\cal B}_B$
represent cubic regions whose faces $\partial_\alpha {\cal B}_A$ and
$\partial_\beta {\cal B}_B$ are identified.  Let $u^{\cal A}_A$ and
$u^{\cal A}_B$ denote the fields evaluated in the cubic regions ${\cal
  B}_A$ and ${\cal B}_B$ respectively.  The required interface
boundary conditions can then be written as,
\begin{eqnarray}
u_B^{\cal B} = {\cal J}^{\,\cal B}{}_{\!\!\cal A}\,
u_A^{\cal A},
\label{e:EllipticContinuityBC}
\end{eqnarray}
to be imposed on the boundary face $\partial_\beta {\cal B}_B$, and
the equation,
\begin{eqnarray}
n^i_A\nabla_{Ai}u_A^{\cal A} &=& n_A^i
J^{*B\beta k}_{A\alpha i}
{\cal J}^{\,{\cal A}}{}_{{\cal B}}\,\nabla_{Bk}u^{\cal B}_B,
\label{e:EllipticSmoothBC}
\end{eqnarray}
to be imposed on the boundary face $\partial_\alpha {\cal B}_A$.  

The required continuity conditions can be imposed numerically by
replacing the elliptic system, Eq.~(\ref{e:EllipticSystem}), with the
equation for the continuity of the fields on the grid points of one of
the boundary faces, $\partial_\beta {\cal B}_B$, and the equation for
the continuity of the normal derivatives on the grid points of the
other face $\partial_\alpha {\cal B}_A$.  Together these boundary
conditions ensure that the global solution to
Eq.~(\ref{e:EllipticSystem}) will have the required continuity and
differentiability at interface boundaries.  Second-order
strongly-elliptic systems can be solved using either Dirichlet or
Neumann type boundary conditions.  Thus the continuity conditions
imposed here are exactly those needed to ensure the well-posedness of
the boundary value problem within each cubic region.

Boundary conditions of this type are already used successfully and
routinely in elliptic-solver codes that implement traditional
multi-patch methods (see e.g. Ref.~\cite{Pfeiffer2003}).  The only
difference between the boundary conditions used in those traditional
multi-patch codes and the ones introduced here is the form of the
Jacobian matrices used to transform the components of tensors and
their derivatives at the interfaces between regions.  In traditional
multi-patch methods these Jacobians are just identity matrices,
because in those cases there was always a smooth global coordinate
basis that could be used to represent tensor fields in all
computational subdomains.  In the multi-cube method introduced here,
these Jacobians contain critical information about the differential
topology of the manifold.

\subsection{Interface Boundary Conditions for Hyperbolic Systems}
\label{s:HyperbolicBC}

A first-order symmetric-hyperbolic system of equations for the
dynamical fields $u^{\cal A}$ (assumed here to be a collection of
tensor fields) can be written in the form
\begin{eqnarray}
\partial_t u^{\cal A} + A^{k{\cal A}}{}_{\cal B}(\mathbf{u})\,\nabla_k u^{\cal
  B} = F^{\cal A}(\mathbf{u}),
\end{eqnarray}
where the characteristic matrix, $A^{k{\cal A}}{}_{\cal
  B}(\mathbf{u})$, and source term, $F^{\cal A}(\mathbf{u})$, may
depend on the fields $u^{\cal A}$ but not their derivatives.  The
script indexes ${\scriptstyle {\cal A}}$, ${\scriptstyle {\cal B}}$,
${\scriptstyle {\cal C}}$, ... in these expressions label the
components of the collection of tensor fields that make up $u^{\cal
  A}$. These systems are called symmetric because, by assumption,
there exists a positive definite metric on the space of fields,
$S_{\cal AB}$, that can be used to transform the characteristic matrix
into a symmetric form: $S_{\cal AC}A^{k\,{\cal C}}{}_{\cal B}\equiv
A^k_{\cal AB}=A^k_{\cal BA}$.

Boundary conditions for symmetric-hyperbolic systems must be imposed
on the incoming characteristic fields of the system.  The
characteristic fields $\hat u^{{\cal K}}$(whose index ${\scriptstyle
  {\cal K}}$ labels the collection of characteristic fields) are
projections of the dynamical fields $u^{\cal A}$ onto the left
eigenvectors of the characteristic matrix (cf. Refs.~\cite{Kidder2005,
  Lindblom2006}),
\begin{eqnarray}
\hat u^{{\cal K}} = e^{{\cal K}}{}_{\!{\cal A}}(\mathbf{n})\, u^{\cal A},
\end{eqnarray}
defined by the equation,
\begin{eqnarray}
e^{{\cal K}}{}_{\!{\cal A}}(\mathbf{n})\,n_kA^{k\,{\cal A}}{}_{\cal B}(u)
= v_{({\cal K})}\,e^{{\cal K}}{}_{\!{\cal B}}(\mathbf{n}).
\end{eqnarray}
The co-vector $n_k$ that appears in this definition is the outward
pointing unit normal to the surface on which the characteristic fields
are evaluated. The eigenvalues $v_{({\cal K})}$ are often referred to
as the characteristic speeds of the system.  The characteristic fields
$\hat u^{{\cal K}}$ represent the independent dynamical degrees of
freedom at the boundaries.  These characteristic fields propagate at
the speeds $v_{({\cal K})}$ (in the short wavelength limit), so
boundary conditions must be given for each incoming characteristic
field, i.e., for each field with speed $v_{({\cal K})}<0$.  No
boundary condition is required (or allowed) for outgoing
characteristic fields, i.e., for any field with $v_{{\cal K})}\geq0$.

The boundary conditions on the dynamical fields $u^{\cal A}$ that
ensure the equations are satisfied across the faces of adjoining cubic
regions are quite simple: data for the incoming characteristic fields
at the boundary of one region are supplied by the outgoing
characteristic fields from the neighboring region.  The boundary
conditions at an interface between cubic regions require that the
dynamical fields $u^{\cal A}_A$ in region ${\cal B}_A$ be transformed
into the tensor basis used in the neighboring region ${\cal B}_B$.
When the dynamical fields $u^{\cal A}$ are a collection of tensor
fields (as assumed here) their components are transformed from one
coordinate representation to another using the Jacobian of the
transformation as described in Eq.~(\ref{e:FieldTransformation}).  In
this case the needed boundary conditions can be stated precisely for
hyperbolic evolution problems: Consider two cubic regions ${\cal B}_A$
and ${\cal B}_B$ whose boundaries $\partial_\alpha {\cal B}_A$ and
$\partial_\beta {\cal B}_B$ are identified by the map $\Psi^{\alpha
  A}_{\beta B}$ as defined in Eq.~(\ref{e:CoordinateMap}).  The
required boundary conditions on the dynamical fields $u^{\cal A}_A$
consist of fixing the incoming characteristic fields $\hat u^{{\cal
    K}}_A$, i.e., those with speeds $v_{({{\cal K}})}<0$, at the
boundary $\partial_\alpha {\cal B}_A$ with data, $u^{\cal B}_B$, from
the fields on the neighboring boundary $\partial_\beta {\cal B}_B$:
\begin{eqnarray}
\hat u^{{\cal K}}_{{A}} &=& e^{{\cal K}}{}_{\!\cal A}(\mathbf{n})
{\cal J}^{\cal A}{}_{\cal B}\,u^{\cal B}_B.
\end{eqnarray}
The matrix of eigenvectors, $e^{{\cal K}}{}_{\!\cal A}(\mathbf{n})$,
that appears in this expression is to be evaluated using the fields
from region ${\cal B}_B$ that have been transformed into region ${\cal
  B}_A$ where the boundary condition is to be imposed.  This boundary
condition must be applied to each incoming characteristic field on
each internal cube face, i.e., on each face that is identified with
the face of a neighboring region.

This type of boundary condition is used routinely and successfully by
hyperbolic evolution codes, such as the Caltech/Cornell SpEC code,
that implement traditional multi-patch methods.  Those traditional
applications differ from the multi-cube methods discussed here only in
the fact that tensors in those traditional cases could always be
expressed in terms of the global coordinate basis.  The generalized
Jacobians ${\cal J}^{\cal A}{}_{\cal B}$ needed to transform tensors
across interface boundaries in those traditional applications of
multi-patch methods are therefore just the identity map.  In the more
general multi-cube construction introduced in
Secs.~\ref{s:TopologicalStructure} and \ref{s:DifferentialStructure},
the Jacobians contain critical information about the differential
topology of the manifold, so the transformations used here must be
slightly more complicated than those used in the traditional
multi-patch case.  Other than that simple difference, however, the
boundary conditions introduced here are the same as those used in the
traditional multi-patch methods.

In some cases, like systems representing second-order tensor wave
equations, the dynamical fields will include a collection of primary
tensor fields plus a collection of secondary fields representing the
first derivatives of the primary fields.  In most cases the secondary
fields can be defined using a covariant derivative, thus making them
tensor fields as well.  The Einstein equations are somewhat
problematic, because the most natural covariant derivative of the
metric tensor (the primary tensor field in this case) vanishes
identically.  Thus first-order symmetric-hyperbolic representations of
the Einstein equations are not generally
co-variant~\cite{Lindblom2006}.  They can be made fully covariant
however by defining the secondary dynamical fields using the covariant
derivative associated with the non-dynamical reference metric that
defines the differential topology of the manifold.  This type of fully
covariant first-order representation of the Einstein system will be
discussed in detail in a future publication.


\section{Numerical Tests of a Multi-Cube Elliptic Equation Solver}
\label{s:TestsEllipticEquations}

This section discusses a series of tests of the numerical solution of
elliptic equations on compact three-manifolds using the multi-cube
methods described in Secs.~\ref{s:TopologicalStructure},
\ref{s:DifferentialStructure}, and \ref{s:BoundaryConditions}.  These
tests find numerical solutions to the equation
\begin{eqnarray}
\nabla^i\nabla_i \psi-c^2 \psi = f,
\label{e:GeneralEllipticEquation}
\end{eqnarray}
where $\psi$ is a scalar field, $\nabla_i$ represents the covariant
derivative associated with a fixed smooth positive-definite metric
$g_{ij}$ on a particular three-manifold, $c$ is a constant, and $f$ is
a fixed source function.  The constant term, with $c^2>0$, ensures the
solution to this equation is unique on any compact three-manifold.
This equation is solved here on the three-manifolds whose multi-cube
representations are described in \ref{s:ExampleRepresentations}: $T^3$
with a flat metric, $S^2\times S^1$ with a round constant-curvature
metric, and $S^3$ with the standard round constant-curvature metric.
The source functions $f$ for these tests are chosen to ensure that the
solutions $\psi$ are non-trivial functions which are known
analytically.  

The accuracy and effectiveness of the numerical solutions of
Eq.~(\ref{e:GeneralEllipticEquation}) are evaluated in two ways.  The
first accuracy indicator used here is the residual, $R$, which
measures how well the numerical solutions satisfy the discrete form of
the differential equations.  This numerical residual is defined as
\begin{eqnarray}
R=\nabla^i\nabla_i \psi_N-c^2 \psi_N - f,
\end{eqnarray}
where $\psi_N$ is the numerical solution of the discrete form of
Eq.~(\ref{e:GeneralEllipticEquation}).  The size of this residual is
monitored for each numerical solution by evaluating its $L^2$ norm 
and computing the normalized residual error quantity, ${\cal E}_R$,
defined as
\begin{eqnarray}
{\cal E}_R = \sqrt{\frac{\int R^2 \sqrt{g}\, d^{\,3}x}
{\int f^2\sqrt{g}\,d^{\,3}x} }.
\label{e:RErrorMeasure}
\end{eqnarray}
The second accuracy indicator used here measures the error in the
numerical solution itself: $\Delta \psi = \psi_E- \psi_N$, where
$\psi_E$ and $\psi_N$ represent the exact analytical solution and the
discrete numerical solutions respectively.  The magnitude of
$\Delta\psi$ is evaluated using the scale invariant $L^2$ measure of
the solution error:
\begin{eqnarray}
{\cal E}_\psi=\sqrt{\frac{\int(\Delta\psi)^2\sqrt{g}\,d^{\,3}x}
                   {\int \psi^2_E\sqrt{g}\,d^{\,3}x} }.
\label{e:PsiErrorMeasure}
\end{eqnarray}

The numerical tests described here were performed using the elliptic
equation solver that is part of the SpEC code~\cite{Pfeiffer2003}.
This code, developed originally by the Caltech/Cornell numerical
relativity collaboration, uses pseudo-spectral methods to represent
functions and evaluate their spatial derivatives.  It solves elliptic
equations using the PETSc toolkit of linear and non-linear equation
solvers.  Each cubic region in the tests described here is subdivided
into one or more computational subregions, on which field components
are represented using Chebyshev basis functions at the Gauss-Lobatto
collocation points.  The structure of these subregions was chosen to
achieve fairly uniform spatial resolution.  The particular choice of
subregions is described in the discussion of each test.

These numerical tests verify that several new ideas introduced in
Secs.~\ref{s:TopologicalStructure}, \ref{s:DifferentialStructure},
\ref{s:BoundaryConditions} and \ref{s:ExampleRepresentations} are
correct, and that these ideas have been implemented correctly in the
SpEC code.  The most fundamental new ideas tested here are the
inter-region boundary conditions, Eqs.~(\ref{e:EllipticContinuityBC})
and (\ref{e:EllipticSmoothBC}), for elliptic equations.  These
internal boundary conditions depend on the Jacobians and their
derivatives, which depend in turn on the inter-region boundary maps in
a critical way for manifolds with non-trivial topologies.  These
Jacobian terms contribute to the boundary conditions in a non-trivial
way even for the simple scalar elliptic
equation~(\ref{e:GeneralEllipticEquation}) used in these tests.  These
tests also depend in a non-trivial way on the multi-cube
representations of the reference metrics
Eqs.~(\ref{e:S2xS1CartesianRoundMetric}) and
(\ref{e:S^3CartesianRoundMetric}) and their associated covariant
derivatives on the manifolds $S^2\times S^1$ and $S^3$.  If any of
these new elements of the multi-cube method were incorrect (or were
implemented incorrectly in the code) the numerical tests described
here would not achieve the exponential convergence in the solution
error measure ${\cal E}_\psi$ that is seen in these tests.

\subsection{Tests of a Multi-Cube Elliptic Equation Solver on $T^3$}

The numerical tests described here use the multi-cube representation
of the three-manifold with topology $T^3$ given in \ref{s:ExampleT3}.
The reference metric in this case is the flat Euclidean metric,
Eq.~(\ref{e:FlatMetricT3}), so the covariant derivatives which appear
in the elliptic Eq.~(\ref{e:GeneralEllipticEquation}) are just the
Cartesian coordinate partial derivatives.  When written in terms of
the multi-cube Cartesian coordinates on $T^3$, therefore, this
equation takes the simple form,
\begin{eqnarray}
\nabla^i\nabla_i\psi - c^2\psi =
\partial_x^2 \psi + \partial_y^2 \psi + \partial_z^2 \psi - c^2 \psi= f.
\label{e:EllipticT3Eq}
\end{eqnarray}
This equation is solved numerically in these tests using the source
function $f$ given by,

\begin{eqnarray}
f(x,y,z)= -(\omega^2+c^2)
\cos\left[\frac{2\pi}{L}\left(k x + \ell y + m z\right)
\right],
\label{e:EllipticSourceT3}
\end{eqnarray}
where $k$, $\ell$, and $m$ are integers, $c$ is a constant $c=1/L$, and
$\omega$ is given by
\begin{eqnarray}
\omega^2 = \left(\frac{2\pi}{L}\right)^2\left(k^2 + \ell^2 + m^2\right).
\label{e:T3EllipticEigenvalue}
\end{eqnarray}
The exact analytical solution to this equation is given by
\begin{eqnarray}
\psi_E(x,y,z) = 
\cos\left[\frac{2\pi}{L} \left(k x + \ell y + m z\right)
\right]. 
\end{eqnarray}

\begin{figure}[ht]
\centerline{\includegraphics[width=2.5in]{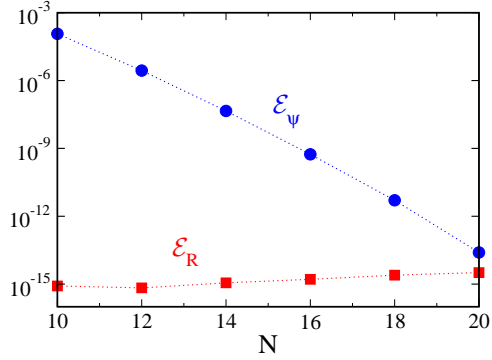}} 
\caption{\label{f:T3EllipticErrorNorms} Errors in the numerical
  solutions $\Delta\psi$ of the elliptic Eq.~(\ref{e:EllipticT3Eq}) on
  $T^3$ with $k=\ell=m=2$, as quantified by the error measures ${\cal
    E}_\psi$ and ${\cal E}_R$.  The parameter $N$ is the number of
  collocation points used for these tests in each spatial direction in
  each computational subregion.}
\end{figure}

The numerical tests of the solutions to
Eqs.~(\ref{e:EllipticT3Eq})--(\ref{e:T3EllipticEigenvalue}) were
performed using a source function with $k=\ell=m=2$.  These tests were
performed on a set of eight computational subregions using a range of
numerical resolutions having $N=8$, $10$, $12$, $14$, $16$, $18$ and
$20$ collocation points respectively in each spatial direction in each
subregion.  These subregions divide the one cubic region ${\cal B}_1$
needed to represent $T^3$ into eight cubes: each half the size of the
region in each spatial direction.  The internal boundary maps between
these subregions are just the trivial identity maps.  The graphs of
the solution errors ${\cal E}_\psi$ and the residual errors ${\cal
  E_R}$, as defined in Eqs.~(\ref{e:RErrorMeasure}) and
(\ref{e:PsiErrorMeasure}), for these tests are shown in
Fig.~\ref{f:T3EllipticErrorNorms}.  The elliptic sover for these
tests were run until the residual errors 
${\cal E_R}$ were reduced to the level of numerical roundoff.
These results demonstrate that the
boundary conditions introduced here on region boundaries were
implemented correctly and efficiently: successfully achieving the
exponential convergence expected of spectral numerical methods.

\subsection{Tests of a Multi-Cube Elliptic Equation Solver on $S^2\times S^1$}

The numerical tests described here use the multi-cube representation
of the three-manifold with topology $S^2\times S^1$ given in
\ref{s:ExampleS2xS1}.  The reference metric used in this case is the
constant-curvature round metric given in terms of angular coordinates
$\{\chi,\theta,\varphi\}$ in Eq.~(\ref{e:S2xS1RoundMetric}), and in
the multi-cube Cartesian coordinates used in these tests in
Eq.~(\ref{e:S2xS1CartesianRoundMetric}).  This choice of reference
metric makes the elliptic Eq.~(\ref{e:GeneralEllipticEquation})
somewhat more complicated in this case.  In terms of the standard
angular coordinates this equation has the form
\begin{eqnarray}
\nabla^i\nabla_i\psi - c^2\psi =
\frac{\partial_\chi^2 \psi}{R_1^2} +
\frac{\partial_\theta
\left[\sin\theta\partial_\theta \psi\right]}{R_2^2\sin\theta}
+ \frac{\partial_\varphi^2 \psi}{R_2^2\sin^2\theta}-c^2\psi= f.
\label{e:EllipticS2xS1Eq}
\end{eqnarray}
This equation is solved numerically in these tests with a source
function $f$ given by,
\begin{eqnarray}
&&
f(\chi,\theta,\varphi)
=-(\omega^2+c^2)\Re\left[e^{i k \chi} Y_{\ell m}(\theta,\varphi)\right],
\label{e:EllipticS2xS1Source}
\end{eqnarray}
where $Y_{\ell m}(\theta,\varphi)$ is the standard $S^2$ spherical
harmonic function, $k$, $\ell$, and $m$ are integers, $c$ is a
constant $c=1/R_2$, $\omega$ is given by
\begin{eqnarray}
\omega^2 = \frac{\ell(\ell+1)}{R_2^2}+\frac{k^2}{R_1^2},
\label{e:EllipticS2xS1Eigenvalue}
\end{eqnarray}
and $\Re[Q]$ denotes the real part of a quantity $Q$.  The exact analytical
solution to this equation is given by
\begin{eqnarray}
\psi_E(\chi,\theta,\varphi) &=& \Re\left[e^{i k \chi} 
Y_{\ell m}(\theta,\varphi)\right].
\label{e:EllipticTestSolS2xS1}
\end{eqnarray}

The numerical solution to this equation is carried out using the
Cartesian coordinates of the multi-cube description of $S^2\times S^1$
described in \ref{s:ExampleS2xS1}.  The covariant derivatives used by
the SpEC code for this test are evaluated using the Cartesian
coordinate representation of the round metric given in
Eq.~(\ref{e:S2xS1CartesianRoundMetric}).  The source function $f$ that
appears on the right side of Eq.~(\ref{e:EllipticS2xS1Eq}), is
evaluated in the multi-cube Cartesian coordinates used for these tests
with the transformations between the angular and Cartesian
coordinates given in Tables~\ref{t:TableIV} and \ref{t:TableV}.

\begin{figure}[ht]
\centerline{\includegraphics[width=2.5in]{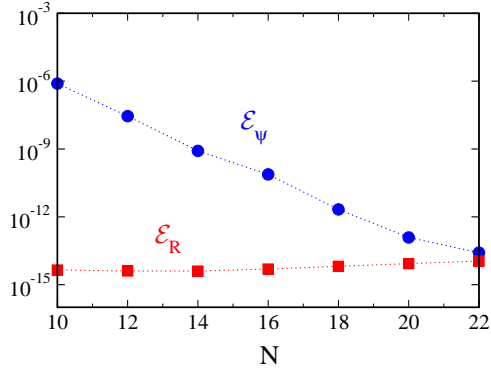}} 
\caption{\label{f:S2xS1EllipticErrorNorms} Errors in the numerical
  solutions $\Delta\psi$ of the elliptic Eq.~(\ref{e:EllipticS2xS1Eq})
  on $S^2\times S^1$ with $k=\ell=m=2$, as quantified by the error
  measures ${\cal E}_\psi$ and ${\cal E}_R$. The parameter $N$ is the number of
  collocation points used for these tests in each spatial direction in
  each computational subregion.}
\end{figure}

The tests performed here used the source function given in
Eqs.~(\ref{e:EllipticS2xS1Source})--(\ref{e:EllipticS2xS1Eigenvalue})
with $k=\ell=m=2$.  These tests used a set of twelve computational
subregions to represent the six cubic regions of $S^2\times S^1$,
cf. Fig.~\ref{f:s2_x_s1}.  These subregions divide each region in
the periodically identified $z$ direction into two subregions.  These
tests were performed using $N=8$, $10$, $12$, $14$, $16$, $18$, $20$
and $22$ collocation points respectively in each spatial direction in
each of the computational subregions.  The boundary conditions at the
inter-region boundaries are based on the maps specified in
Table~\ref{t:TableIII}.  The graphs of the solution errors ${\cal
  E}_\psi$ and the residual errors ${\cal E_R}$, as defined in
Eqs.~(\ref{e:RErrorMeasure}) and (\ref{e:PsiErrorMeasure}), for these
tests are shown in Fig.~\ref{f:S2xS1EllipticErrorNorms}.  
The elliptic sover for these
tests were run until the residual errors 
${\cal E_R}$ were reduced to the level of numerical roundoff.
This graph
demonstrates, for the non-trivial $S^2\times S^1$ case, that the
computational region boundary conditions developed here have been
implemented correctly and efficiently, achieving the exponential
convergence expected of spectral numerical methods.

\subsection{Tests of a Multi-Cube Elliptic Equation Solver on $S^3$}

The numerical tests described here use the multi-cube representation
of the three-manifold with topology $S^3$ given in \ref{s:ExampleS3}.
The reference metric used in this case is the standard
constant-curvature round metric for $S^3$ given in terms of angular
coordinates $\{\chi,\theta,\varphi\}$ in Eq.~(\ref{e:S^3RoundMetric}),
and in the multi-cube Cartesian coordinates used in these tests in
Eq.~(\ref{e:S^3CartesianRoundMetric}).  This choice of reference
metric fixes the elliptic Eq.~(\ref{e:GeneralEllipticEquation}) to
have the form, 
\begin{eqnarray}
\nabla^i\nabla_i\psi - c^2\psi =
\frac{\partial_\chi\left[\sin^2\chi\partial_\chi \psi\right]}{R_3^2\sin^2\chi}
+\frac{\partial_\theta\left[\sin\theta\partial_\theta \psi\right]}
{R_3^2\sin\theta\sin^2\chi}
+\frac{\partial_\varphi^{\,2}\,\psi}{R_3^2\sin^2\theta\sin^2\chi}
-c^2 \psi=f,
\label{e:EllipticS3Eq}
\end{eqnarray}
when expressed in terms of the standard angular coordinates
$\{\chi,\theta,\varphi\}$ used on $S^3$.  The source function $f$ used
in these numerical tests is given by,
\begin{eqnarray}
f(\chi,\theta,\varphi) =-(\omega^2+c^2) 
\Re\left[Y_{k\ell m}(\chi,\theta,\varphi)\right],
\label{e:EllipticS3Source}
\end{eqnarray}
where the $Y_{k\ell m}(\chi,\theta,\varphi)$ are the $S^3$ spherical
harmonics described in \ref{s:HyperSphericalHarmonics}, $k$, $\ell$,
and $m$ are integers, $c$ is a constant $c=1/R_3$, and $\omega$ is
given by
\begin{eqnarray}
\omega^2 = \frac{k(k+2)}{R_3^2}.
\label{e:EllipticS3Eigenvalues}
\end{eqnarray}
The exact analytical solution to this equation is given by
\begin{eqnarray}
\psi_E(\chi,\theta,\varphi) &=& \Re\left[
Y_{k\ell m}(\chi,\theta,\varphi)\right].
\label{e:EllipticTestSolS3}
\end{eqnarray}

The numerical solutions of Eq.~(\ref{e:EllipticS3Eq}) are carried out
for these tests using the multi-cube representation of $S^3$ described
in \ref{s:ExampleS3}.  The covariant derivatives used by the SpEC code
for this test are evaluated using the multi-cube Cartesian coordinate
representation of the round metric on $S^3$ given in
Eq.~(\ref{e:S^3CartesianRoundMetric}).  The source function $f$,
defined in Eq.~(\ref{e:EllipticS3Source}), is evaluated in terms of
the multi-cube Cartesian coordinates for these tests using the
transformations between the angular and the Cartesian coordinates
given in Tables~\ref{t:TableVIII} and \ref{t:TableIX}.

\begin{figure}[h]
\centerline{\includegraphics[width=2.5in]{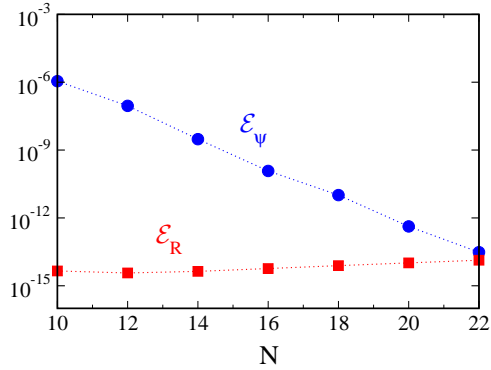}} 
\caption{\label{f:S3EllipticErrorNorms} Errors in the numerical
  solutions $\Delta\psi$ of the elliptic
  Eq.~(\ref{e:EllipticS3Eq}) on $S^3$ with $k=\ell=m=2$ as quantified by
  the error measures ${\cal E}_\psi$ and ${\cal E}_R$. The parameter $N$
  is the number of collocation points used for these tests in each
  spatial direction in each computational subregion.}
\end{figure}
The numerical tests described here solved the elliptic
Eqs.~(\ref{e:EllipticS3Eq})--(\ref{e:EllipticS3Eigenvalues}) with the
parameter values $k=\ell=m=2$ in the source function $f$.  These tests
were done using a set of eight computational subregions, corresponding
to the eight cubic regions needed to represent $S^3$,
cf. Fig.~\ref{f:s3}.  These tests used $N=8$, $10$, $12$, $14$, $16$,
$18$, $20$ and $22$ collocation points respectively in each spatial
direction in each of the computational subregions.  The boundary
conditions at the region boundaries for these tests are based on the
interface identification maps specified in Table~\ref{t:TableVIII}.  The
graphs of the solution errors ${\cal E}_\psi$ and the residual errors
${\cal E_R}$, defined in Eqs.~(\ref{e:RErrorMeasure}) and
(\ref{e:PsiErrorMeasure}), for these tests are shown in
Fig.~\ref{f:S3EllipticErrorNorms}.  
The elliptic sover for these
tests were run until the residual errors 
${\cal E_R}$ were reduced to the level of numerical roundoff.
This graph demonstrates for
another non-trivial example that the inter-region boundary conditions
developed here have been implemented correctly and efficiently.
Figure~\ref{f:S3EllipticErrorNorms} also demonstrates that these
numerical tests have achieved the exponential convergence expected of
spectral numerical methods.

\section{Numerical Tests of a Multi-Cube Hyperbolic Equation Solver}
\label{s:TestsHyperbolicEquations}

This section discusses numerical tests of the multi-cube methods for
solving hyperbolic evolution equations on compact three-manifolds as
described in Secs.~\ref{s:TopologicalStructure},
\ref{s:DifferentialStructure}, and \ref{s:BoundaryConditions}.  These
tests find numerical solutions to the scalar wave equation
\begin{eqnarray}
-\partial_t^{\,2} \psi + \nabla^i\nabla_i \psi=0,
\label{e:GeneralWaveEquation}
\end{eqnarray}
where $\nabla_i$ represents the spatial covariant derivative on the
fixed geometry of the spatial three-manifold.  This equation is solved
here on the three-manifolds described in
\ref{s:ExampleRepresentations}: $T^3$ with a flat metric, $S^2\times
S^1$ with the constant curvature round metric, and $S^3$ with the
standard constant-curvature round metric.

These wave equations are converted to first-order symmetric-hyperbolic form
before solving them numerically.  The list of dynamical fields
$u^\alpha=\{\psi,\Pi,\Phi_i\}$ is therefore expanded to include the
first derivatives of $\psi$: $\Pi=-\partial_t\psi$, and
$\Phi_i=\partial_i\psi$.  Constraint damping is used to enforce the
constraint, 
\begin{eqnarray}
{\cal C}_i\equiv\partial_i\psi-\Phi_i=0,
\end{eqnarray}
using the methods developed in Ref.~\cite{Holst2004} with constraint
damping parameter $\gamma_2=1$.

Exact analytical solutions exist to these wave equations on the
three-manifolds used in these tests.  Therefore the effectiveness and
efficiency of the evolution code can be tested in these cases by
comparing numerical solutions $\psi_N$ to this equation with the known
analytical solutions $\psi_E$.  The accuracy, and convergence
properties, of the code can be measured therefore by monitoring the
$L^2$ norms of $\Delta \psi = \psi_E- \psi_N$ using the solution error
measure defined in Eq.~(\ref{e:PsiErrorMeasure}).  It is also useful
to monitor the constraint violation errors ${\cal C}_i$.  This is done
by constructing the constraint error measure:
\begin{eqnarray}
{\cal E_C} \equiv \sqrt{\frac{\int g^{ij} {\cal C}_i{\cal C}_j \sqrt{g}\,
d^{\,3}x}
{\int g^{ij}\left(\Phi_i\Phi_j+\partial_i\psi\partial_j\psi\right)
\sqrt{g}\,d^{\,3}x}}.
\end{eqnarray}
This constraint error measure is invariant under changes in the overall scale
of the solution, and to changes in the coordinates used to represent the
solution.

The tests performed here use the scalar wave evolution system that is
implemented as part of the SpEC code~\cite{Holst2004,Scheel2004}.
This code, developed originally by the Caltech/Cornell numerical
relativity collaboration, uses pseudo-spectral methods to evaluate
spatial derivatives, and the method of lines to approximate the
hyperbolic system of partial differential equations as sets of coupled
ordinary differential equations on each collocation point.  These tests
use an eighth order Dormand-Prince~\cite{DormandPrince1980} algorithm
to integrate the method of lines ordinary differential equations in
time.  Each cubic region in these tests is subdivided into one or more
computational subregions, on which field components are represented
using Chebyshev basis functions at the Gauss-Labatto collocation
points.  The structure of these subregions was chosen to achieve
fairly uniform spatial resolution.  The particular choice of
subregions is described in the discussion of each particular test.

\subsection{Tests of a Multi-Cube Hyperbolic Equation Solver on $T^3$}

The numerical tests described here use the multi-cube representation
of the three-manifold with topology $T^3$ given in \ref{s:ExampleT3}.
The reference metric in this case is the flat Euclidean metric,
Eq.~(\ref{e:FlatMetricT3}), so the spatial covariant derivatives which
appear in the wave Eq.~(\ref{e:GeneralWaveEquation}) are just the
Cartesian coordinate partial derivatives.  When written in terms of
the multi-cube Cartesian coordinates on $T^3$, therefore, the wave
equation takes the simple form,
\begin{eqnarray}
-\partial_t^{\,2} \psi + \nabla^i\nabla_i \psi=
-\partial_t^{\,2}\psi + \partial_x^{\,2} \psi + \partial_y^{\,2} \psi 
+ \partial_z^{\,2} \psi=0.
\label{e:T3WaveEquation}
\end{eqnarray}
The idea is to solve this equation numerically with initial data:
\begin{eqnarray}
\psi(t,x,y,z)\left|_{t=0}\right. 
&=& \cos\left[\frac{2\pi}{L}\left(k x + \ell y + m z\right)
\right]
,\label{e:T3InitialDataPsi}
\\
\!\!\!\!\!\!\!\!
\partial_t\psi(t,x,y,z)\left|_{t=0}\right. &=& -\omega \sin\left[
\frac{2\pi}{L}\left(k x + \ell y + m z\right)\right],\quad
\label{e:T3InitialDataPi}
\end{eqnarray}
where $k$, $\ell$, and $m$ are integers, and $\omega$ is given by
\begin{eqnarray}
\omega^2 = \left(\frac{2\pi}{L}\right)^2\left(k^2 + \ell^2 + m^2\right).
\end{eqnarray}
The exact solution to this initial value problem is given analytically by
\begin{eqnarray}
\psi_E(t,x,y,z) = 
\cos\left[\omega t+\frac{2\pi}{L} \left(k x + \ell y + m z\right)
\right]. 
\end{eqnarray}

\begin{figure}[ht]
\begin{picture}(0,115)(0,75)
\put(5,70){
\includegraphics[width=2.5in]{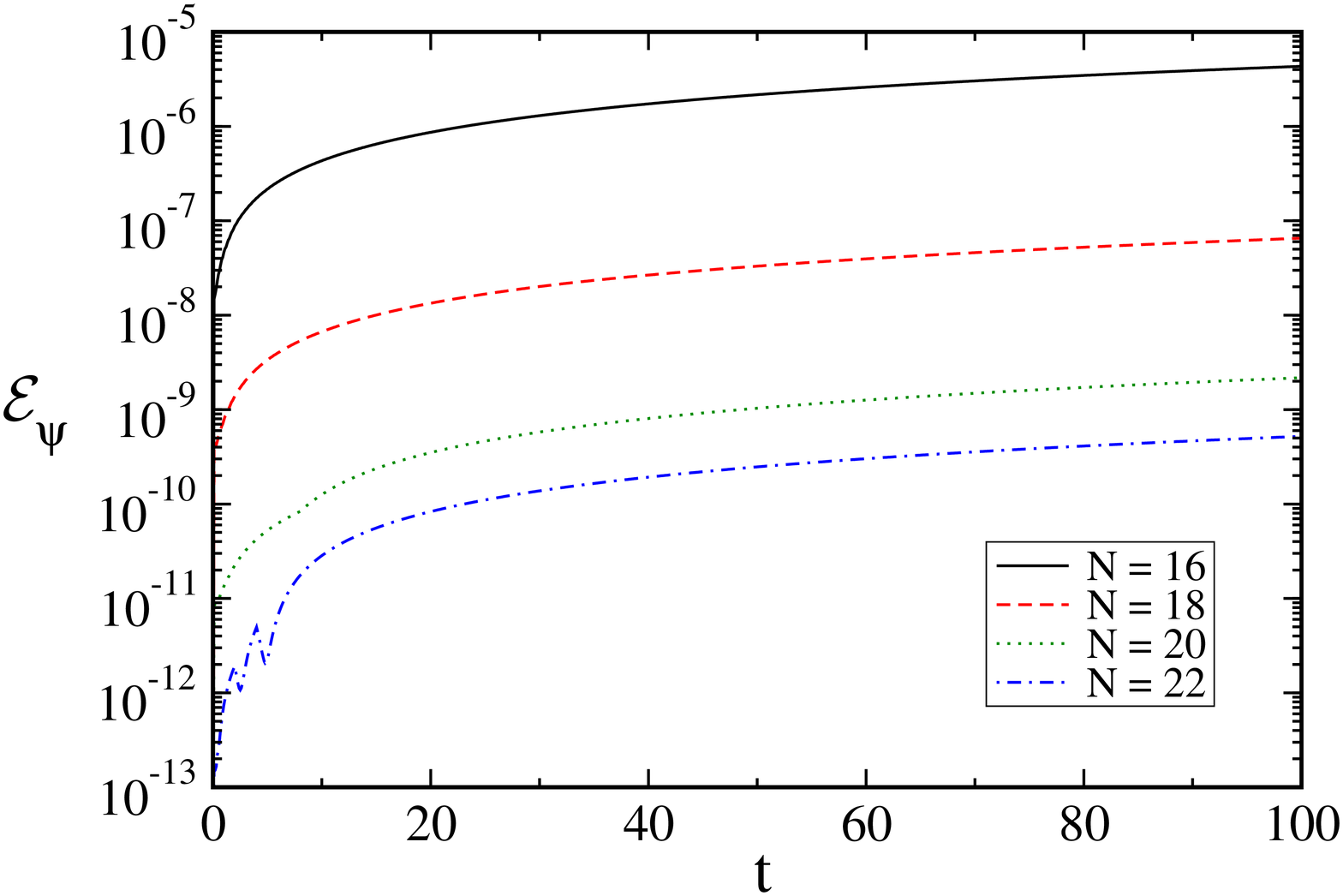}
} 
\put(195,70){
\includegraphics[width=2.5in]{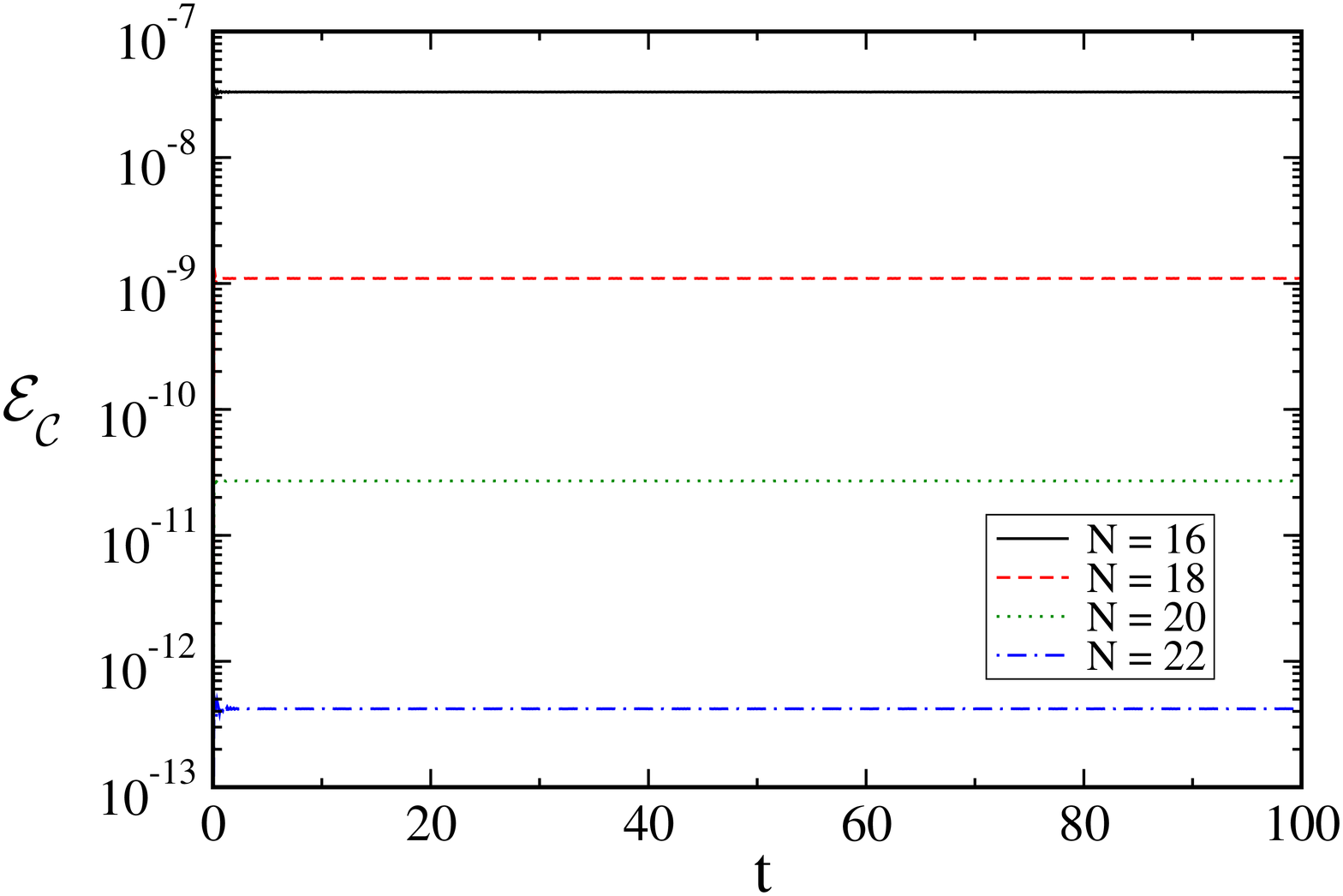}
}
\end{picture}
\caption{\label{f:T3ErrorNorms} Left: Errors in the numerical solutions
$\Delta\psi$ for the $T^3$ evolutions with $k=\ell=m=2$
as measured by the quantity ${\cal E}_\psi$.
Right: Constraint errors ${\cal C}_i$
for the $T^3$ evolutions with $k=\ell=m=2$
as measured by the quantity ${\cal E_C}$.}
\end{figure}

The numerical solution of the wave Eq.~(\ref{e:T3WaveEquation}) for
these tests was performed on a set of eight computational subregions.
These subregions divide the one cubic region needed to represent $T^3$
into eight cubes, each half the size of the region in each spatial
direction.  The internal boundary maps between these subregions are
just the trivial identity maps.  These hyperbolic evolution tests were
performed using the initial data given in
Eqs.~(\ref{e:T3InitialDataPsi}) and (\ref{e:T3InitialDataPi}) with
$k=\ell=m=2$.  These tests used computational subregions having
$N=16$, $18$, $20$ and $22$ collocation points respectively in each
spatial direction.  The graphs of the solution errors ${\cal E}_\psi$
and the constraint violation errors ${\cal E_C}$ for these tests are
shown in Fig.~\ref{f:T3ErrorNorms}.  These graphs demonstrate that the
numerical methods described here successfully achieve the exponential
convergence expected of spectral numerical methods.  The slow growth
in time of the solution error ${\cal E}_\psi$, seen in the left side
of Fig.~\ref{f:T3ErrorNorms} is linear in time. This type of error is
a common feature of the ordinary differential equation integrator used
for these tests.

\subsection{Tests of a Multi-Cube 
Hyperbolic Equation Solver on $S^2\times S^1$}

The numerical tests described here use the multi-cube representation
of the three-manifold with topology $S^2\times S^1$ given in
\ref{s:ExampleS2xS1}.  The reference metric used in this case is the
constant-curvature round metric given in terms of angular coordinates
$\{\chi,\theta,\varphi\}$ in Eq.~(\ref{e:S2xS1RoundMetric}), and in
the multi-cube Cartesian coordinates used in these tests in
Eq.~(\ref{e:S2xS1CartesianRoundMetric}).  This choice of reference
metric fixes the wave Eq.~(\ref{e:GeneralEllipticEquation}) to have
the form
\begin{eqnarray}
-\partial_t^{\,2} \psi + \nabla^i\nabla_i \psi=
-\partial_t^{\,2}\psi +  \frac{\partial_\chi^{\,2} \psi}{R_1^2} +
\frac{\partial_\theta
\left[\sin\theta\partial_\theta \psi\right]}{R_2^2\sin\theta} 
+ \frac{\partial_\varphi^{\,2} \psi}{R_2^2\sin^2\theta}=0.
\label{e:S2xS1WaveEquation}
\end{eqnarray}
when expressed in terms of the angular coordinates
$\{\chi,\theta,\varphi\}$ used on $S^2\times S^1$.  The idea is to
solve this equation numerically with initial data:
\begin{eqnarray}
\psi(t,\theta,\varphi,\chi)_{t=0} &=& \Re\left[e^{i k \chi} Y_{\ell m}(\theta,\varphi)\right],\label{e:S2xS1InitialDataPsi}\\
\partial_t\psi(t,\theta,\varphi,\chi)_{t=0} &=& 
\Re\left[ i\omega e^{i k \chi} Y_{\ell m}(\theta,\varphi)\right],
\label{e:S2xS1InitialDataPi}
\end{eqnarray}
where $Y_{\ell m}(\theta,\varphi)$ are the standard $S^2$ spherical
  harmonics, $k$, $\ell$, and $m$ are integers, $\omega$ is given by
\begin{eqnarray}
\omega^2 = \frac{\ell(\ell+1)}{R_2^2}+\frac{k^2}{R_1^2},
\end{eqnarray}
and $\Re[Q]$ denotes the real part of the quantity $Q$.  The exact
solution to this initial value problem is given analytically by
\begin{eqnarray}
\psi_E(t,\theta,\varphi,\chi) &=& \Re\left[e^{i\omega t+i k \chi} 
Y_{\ell m}(\theta,\varphi)\right].
\end{eqnarray}

The numerical solution of Eq.~(\ref{e:S2xS1WaveEquation}) is carried
out using the Cartesian coordinates of the multi-cube description of
$S^2\times S^1$ described in \ref{s:ExampleS2xS1}.  The spatial
covariant derivatives used by the SpEC code for this test are
evaluated using the Cartesian coordinate representation of the round
metric given in Eq.~(\ref{e:S2xS1CartesianRoundMetric}).  The initial
data, Eqs.~(\ref{e:S2xS1InitialDataPsi}) and
(\ref{e:S2xS1InitialDataPi}), used for these tests are evaluated in
the multi-cube Cartesian coordinates with the transformations between
the angular and Cartesian coordinates given in
Tables~\ref{t:TableIV} and \ref{t:TableV}.

\begin{figure}[ht]
\begin{picture}(0,115)(0,75)
\put(5,70){
\includegraphics[width=2.5in]{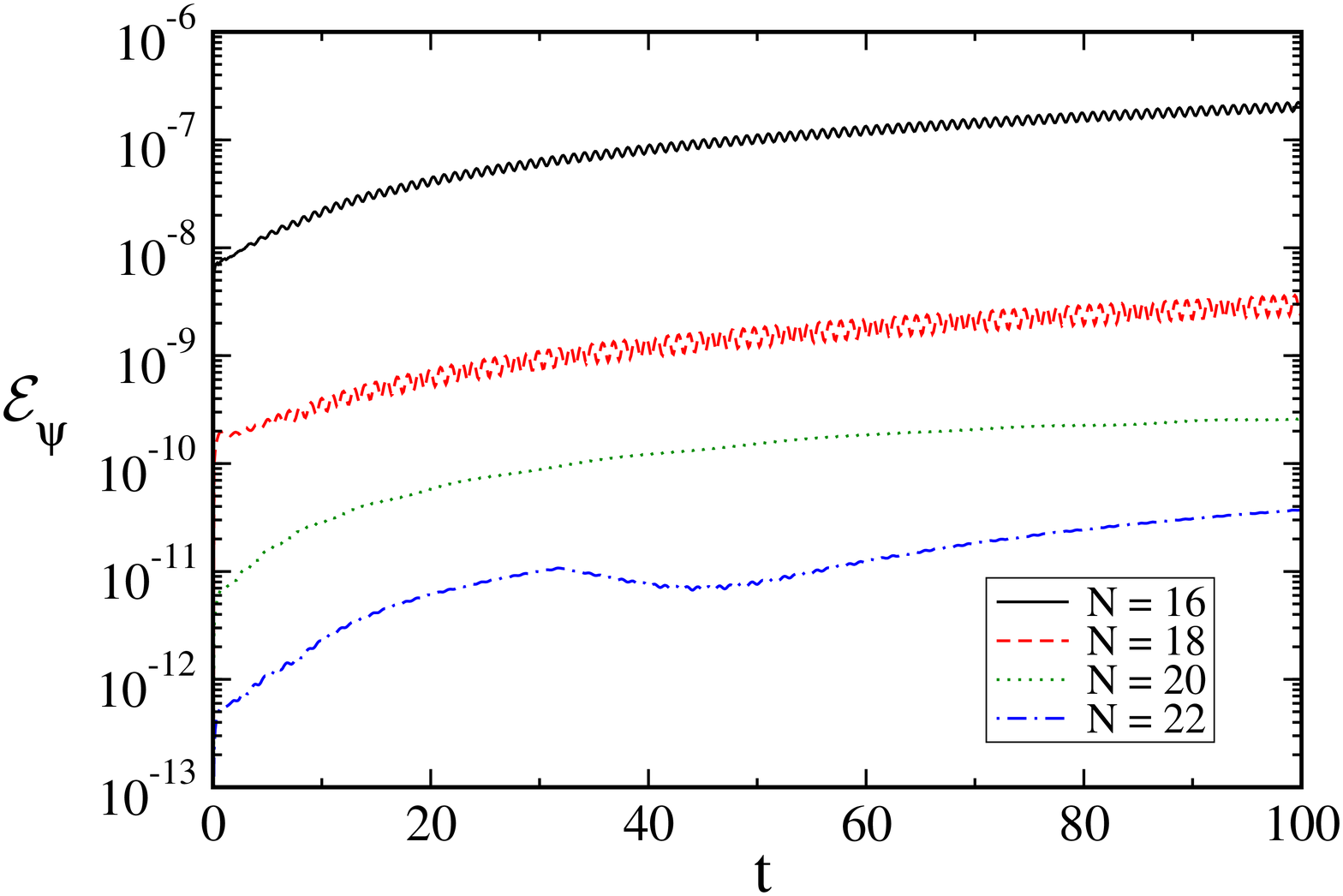}
} 
\put(195,70){
\includegraphics[width=2.5in]{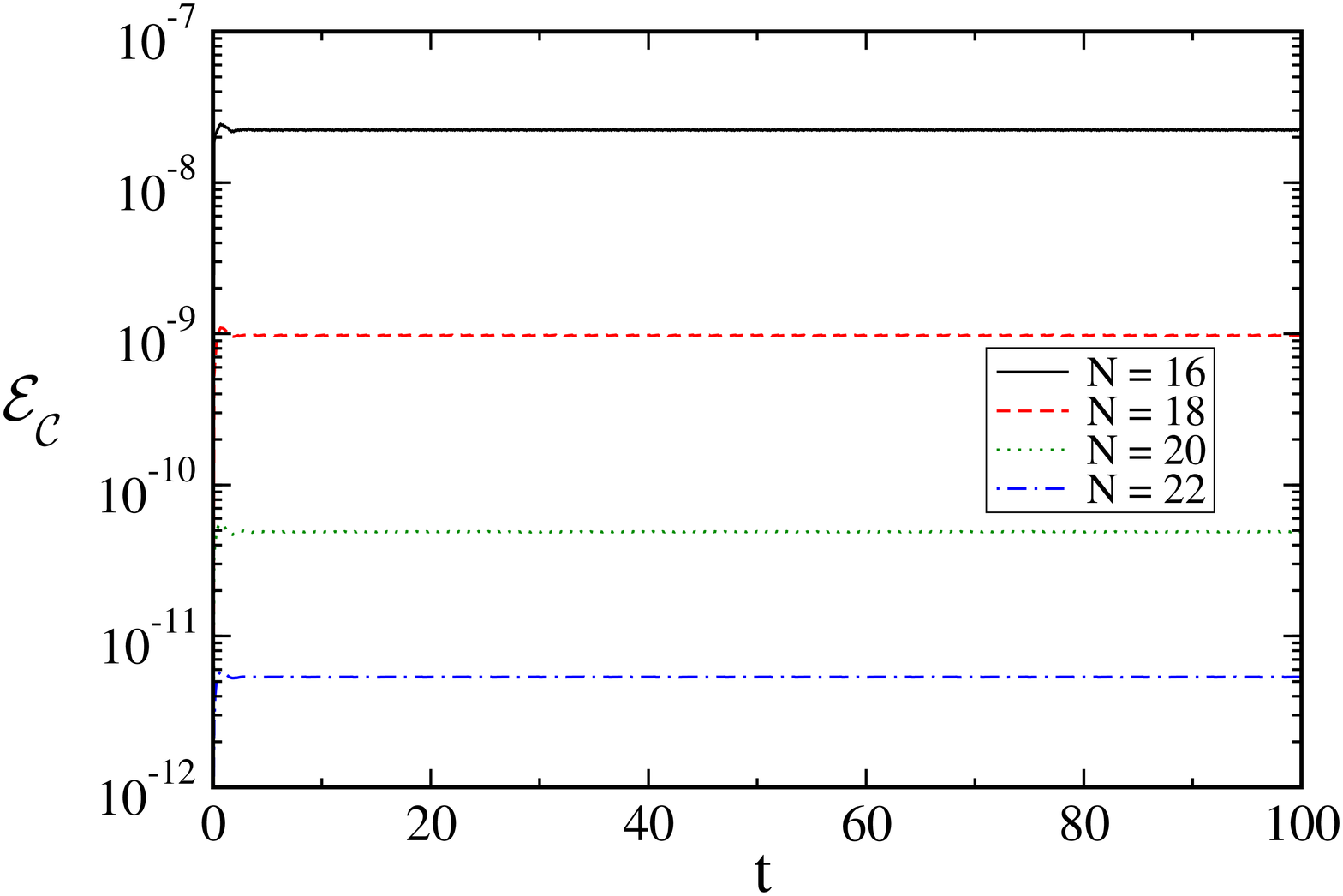}
}
\end{picture}
\caption{\label{f:S2xS1ErrorNorms} Left: Errors in the numerical
  solutions $\Delta\psi$ for the $S^2\times S^1$ evolutions with
  $k=\ell=m=2$ as measured by the quantity ${\cal E}_\psi$.  Right:
  Constraint errors ${\cal C}_i$ for the $S^2\times S^1$ evolutions
  with $k=\ell=m=2$ as measured by the quantity ${\cal E_C}$.}
\end{figure}

The numerical solution of the scalar wave
Eq.~(\ref{e:S2xS1WaveEquation}) for these tests was performed on a
set of twelve computational subregions.  These subregions divide the
six cubic regions needed to represent $S^2\times S^1$,
cf. Fig.~\ref{f:s2_x_s1}, into cubes that are half the size of the
region in the $z$ direction.  The internal boundary maps between these
subregions are just the trivial identity maps, while the maps between
regions are those given in Table~\ref{t:TableIII}.  These hyperbolic
evolution tests were performed using the initial data given in
Eqs.~(\ref{e:S2xS1InitialDataPsi}) and (\ref{e:S2xS1InitialDataPi})
with $k=\ell=m=2$.  These tests were performed on computational
subregions having $N=16$, $18$, $20$ and $22$ collocation points
respectively in each spatial direction.  The graphs of the solution
errors ${\cal E}_\psi$ and the constraint violation errors ${\cal
  E_C}$ for these tests are shown in Fig.~\ref{f:S2xS1ErrorNorms}.
These graphs demonstrate that the numerical methods described here
successfully achieve the exponential convergence expected of spectral 
numerical methods.  The slow growth in time of the solution error
${\cal E}_\psi$, seen in left side of Fig.~\ref{f:S2xS1ErrorNorms} is
(mostly) linear in time. This growth in the error is a common feature
of the ordinary differential equation integrator used for these tests.

\subsection{Tests of a Multi-Cube Hyperbolic Equation Solver on $S^3$}

The numerical tests described here use the multi-cube representation
of the three-manifold with topology $S^3$ given in \ref{s:ExampleS3}.
The reference metric used in this case is the standard
constant-curvature round metric for $S^3$ given in terms of angular
coordinates $\{\chi,\theta,\varphi\}$ in Eq.~(\ref{e:S^3RoundMetric}),
and in the multi-cube Cartesian coordinates used in these tests in
Eq.~(\ref{e:S^3CartesianRoundMetric}).  This choice of reference
metric fixes the wave Eq.~(\ref{e:GeneralWaveEquation}) to
have the form, 
\begin{eqnarray}
-\partial_t^{\,2} \psi + \nabla^i\nabla_i \psi=
-\partial_t^{\,2}\psi + 
\frac{\partial_\chi\left[\sin^2\chi\partial_\chi \psi\right]}{R_3^2\sin^2\chi}
+\frac{\partial_\theta\left[\sin\theta\partial_\theta \psi\right]}
{R_3^2\sin\theta\sin^2\chi}
+\frac{\partial_\varphi^{\,2}\,\psi}{R_3^2\sin^2\theta\sin^2\chi}=0,
\label{e:S3WaveEquation}
\end{eqnarray}
when expressed in terms of the standard angular coordinates
$\{\chi,\theta,\varphi\}$ used on $S^3$.  This equation is solved
numerically with initial data:
\begin{eqnarray}
\psi(t,\theta,\varphi,\chi)_{t=0} &=& 
\Re\left[Y_{k\ell m}(\chi,\theta,\varphi)\right],
\label{e:S3InitialDataPsi}\\ 
\partial_t\psi(t,\theta,\varphi,\chi)_{t=0} &=& 
\Re\left[ i\omega Y_{k\ell m}(\chi,\theta,\varphi)\right],
\label{e:S3InitialDataPi}
\end{eqnarray}
where $Y_{k\ell m}$ is the $S^3$ spherical harmonic function defined in
\ref{s:HyperSphericalHarmonics}, $k$, $\ell$, and $m$ are integers,
and $\omega$ is given by
\begin{eqnarray}
\omega^2 = \frac{k(k+2)}{R_3^2}.
\end{eqnarray}
The solution to this initial
value problem is given analytically by
\begin{eqnarray}
\psi_E(t,\theta,\varphi,\chi) &=& \Re\left[e^{i\omega t} Y_{k\ell
    m}(\chi,\theta,\varphi)\right].
\end{eqnarray}

The numerical solution of Eq.~(\ref{e:S3WaveEquation}) is carried out
using the Cartesian coordinates of the multi-cube description of $S^3$
described in \ref{s:ExampleS3}.  The spatial covariant derivatives
used by the SpEC code for this test are evaluated using the Cartesian
coordinate representation of the round metric given in
Eq.~(\ref{e:S^3CartesianRoundMetric}).  The initial data,
Eqs.~(\ref{e:S3InitialDataPsi}) and (\ref{e:S3InitialDataPi}), used
for these tests are evaluated in the multi-cube Cartesian coordinates
with the transformations between the angular and Cartesian coordinates
given in Table~\ref{t:TableVIII} and \ref{t:TableIX}.

\begin{figure}[ht]
\begin{picture}(0,115)(0,75)
\put(5,70){
\includegraphics[width=2.5in]{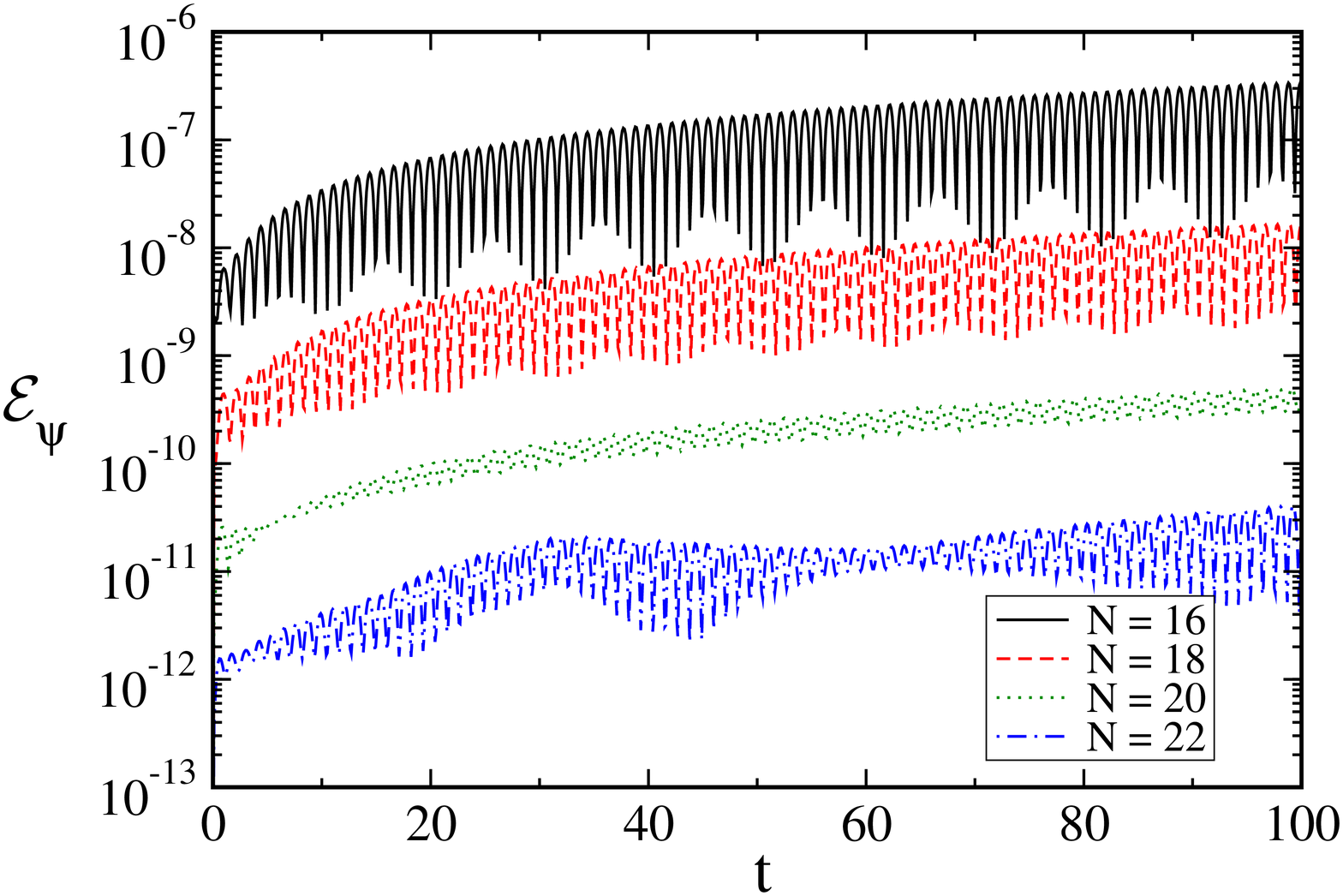}
} 
\put(195,70){
\includegraphics[width=2.5in]{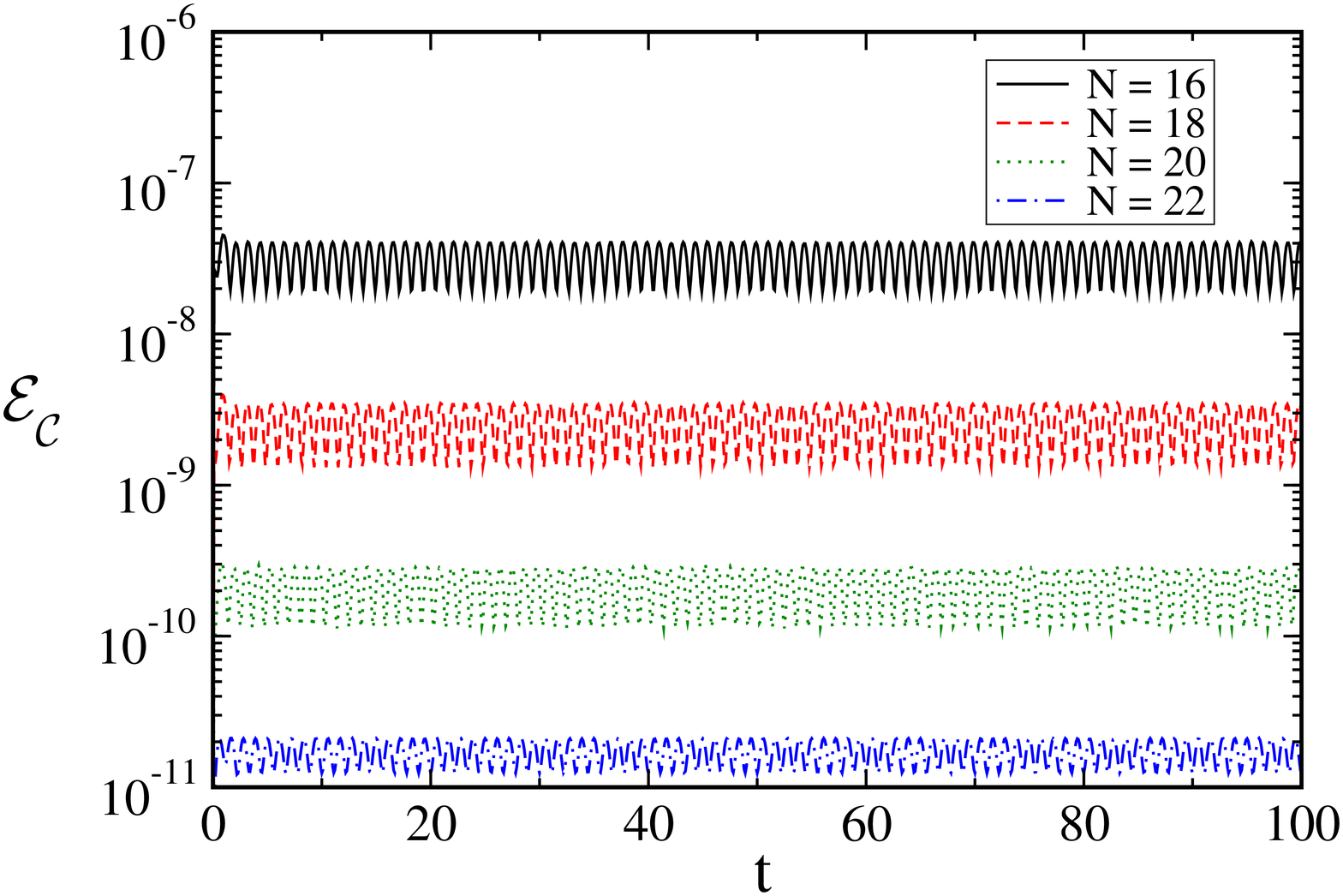}
}
\end{picture}
\caption{\label{f:S3ErrorNorms} Left: Errors in the numerical
  solutions $\Delta\psi$ for the $S^3$ evolutions with
  $k=\ell=m=2$ as measured by the quantity ${\cal E}_\psi$. Right:
  Constraint errors ${\cal C}_i$ for the $S^3$ evolutions with
  $k=\ell=m=2$ as measured by the quantity ${\cal E_C}$.}
\end{figure}

The numerical solution of the scalar wave
Eq.~(\ref{e:S3WaveEquation}) for these tests was performed on a set of
eight computational subregions.  These subregions are identical to the
eight cubic regions needed to represent $S^3$, cf. Fig.~\ref{f:s3}.
The maps between regions are those given in Table~\ref{t:TableVII}.  The
hyperbolic evolution test was performed using the initial data given
in Eqs.~(\ref{e:S3InitialDataPsi}) and (\ref{e:S3InitialDataPi}) with
$k=\ell=m=2$.  These tests were performed on computational subregions
having $N=16$, $18$, $20$ and $22$ collocation points respectively in
each spatial direction.  The graphs of the solution errors ${\cal
  E}_\psi$ and the constraint violation errors ${\cal E_C}$ for these
tests are shown in Fig.~\ref{f:S3ErrorNorms}.  These graphs
demonstrate that the numerical methods described here successfully
achieve the exponential convergence expected of spectral numerical
methods.  The slow growth in time of the solution error ${\cal
  E}_\psi$, seen in the left side of Fig.~\ref{f:S3ErrorNorms} is
(mostly) linear in time. This growth in the error is a common feature
of the ordinary differential equation integrator used for these tests.


\section*{Acknowledgment} 
We thank Michael Holst for helpful discussions about elliptic systems
of equations and about triangulations of topological manifolds, and we
thank Oliver Rinne and Manuel Tiglio for providing a number of useful
comments on a draft of this paper.  Part of this research was
completed while LL was visiting the Max Planck Institute for
Gravitational Physics (Albert Einstein Institute) in Golm, Germany.
This research was supported in part by a grant from the Sherman
Fairchild Foundation, and by NSF grants PHY-1005655, PHY-1068881
and DMS-1065438.

\vfill\break
\appendix
\section{Examples of Multi-Cube Representations of Three-Manifolds}
\label{s:ExampleRepresentations}

This appendix describes the construction of multi-cube representations
of manifolds using the methods developed in
Secs.~\ref{s:TopologicalStructure} and \ref{s:DifferentialStructure}.
Each multi-cube representation consists of a set of non-overlapping
cubes ${\cal B}_A$ that cover the manifold, a set of maps
$\Psi^{A\alpha}_{B\beta}$ that identify the faces of neighboring
cubes, and finally a smooth positive definite reference metric
$g_{ij}$ used to define the differential structure on the manifold.
The construction of these multi-cube structures is described here for
three common three-manifolds: the three-torus $T^3$ with a flat
reference metric, the spherical-torus $S^2\times S^1$ with a
constant-curvature round-sphere metric, and the three-sphere $S^3$
with the standard constant-curvature round-sphere metric.
These examples are used in Secs.~\ref{s:TestsEllipticEquations} and
\ref{s:TestsHyperbolicEquations} to illustrate the solution of partial
differential equations on multi-cube manifolds using the methods
developed in Sec.~\ref{s:BoundaryConditions}.

\subsection{Multi-Cube Representation of $T^3$}
\label{s:ExampleT3}

The simplest example of a multi-cube manifold is the three-torus,
$T^3$.  Only a single cube ${\cal B}_1$ is needed to cover this
manifold, and it is most convenient to locate this cube at the origin
in $R^3$ so $\vec c_1=(0,0,0)$.  Opposite faces of this cube are
identified without rotation or reflection to obtain the $T^3$
topology: $\partial_{+x}{\cal B}_1 \leftrightarrow \partial_{-x} {\cal
  B}_1$, $\partial_{+y}{\cal B}_1 \leftrightarrow \partial_{-y} {\cal
  B}_1$, and $\partial_{+z}{\cal B}_1 \leftrightarrow \partial_{-z}
{\cal B}_1$.  The maps, $\Psi^{1\pm x}_{1\mp x}$, $\Psi^{1\pm y}_{1\mp
  y}$, and $\Psi^{1\pm z}_{1\mp z}$, needed to effect these
identifications are defined by Eq.~(\ref{e:CoordinateMap}) with the
rotation matrices, $C^{A\alpha}_{B\beta}$, being just the identity
matrices: ${\mathbf C}^{1+x}_{1-x}={\mathbf C}^{1+y}_{1-y}={\mathbf
  C}^{1+z}_{1-z}={\mathbf I}$.  The three-torus $T^3$ admits a smooth
flat metric, so a convenient choice of reference metric for this
manifold is:
\begin{eqnarray}
ds^2 = g_{ij}dx^idx^j = dx^2 + dy^2 +dz^2,
\label{e:FlatMetricT3}
\end{eqnarray}
where $x$, $y$ and $z$ are the multi-cube Cartesian coordinates that
label points in ${\cal B}_1$.

\subsection{Multi-Cube Representation of $S^2 \times S^1$}
\label{s:ExampleS2xS1}

\begin{figure}[ht]
\centerline{\includegraphics[width=2.5in]{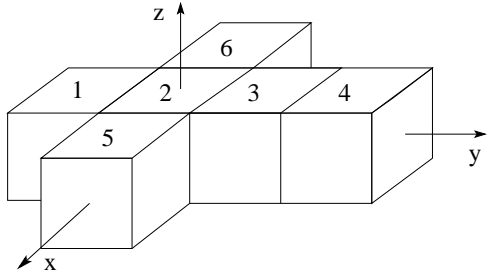}}
\caption{\label{f:s2_x_s1} The three-manifold $S^2\times S^1$ is
  represented using the six cubic regions illustrated here.  The
  faces of these cubes are identified using the maps described
  in Table~\ref{t:TableIII}.  This representation of $S^2\times S^1$ is
  based on the commonly used ``cubed-sphere'' representation of
  $S^2$.}
\end{figure}

The manifold $S^2\times S^1$ can be covered by a set of six cubic
regions ${\cal B}_A$ with ${\scriptstyle A}=\{1, ..., 6\}$.  A
convenient way to arrange these cubes in $R^3$ is illustrated in
Fig.~\ref{f:s2_x_s1}.  The values of the cube-center location vectors
$\vec c_A$ for this configuration is summarized in
Table~\ref{t:TableII}.  The inner faces of the touching cubes in
Fig.~\ref{f:s2_x_s1} are connected by identity maps, while the outer
faces are identified using the maps described by
Eq.~(\ref{e:CoordinateMap}) with the rotation matrices ${\mathbf
  C}^{A\alpha}_{B\beta}$ given in Table~\ref{t:TableIII}.  This
representation of $S^2\times S^1$ is constructed by taking the
Cartesian product of $S^1$ (the periodically identified $z$-axis in
this representation) with the commonly used ``cubed-sphere''
representation of $S^2$~\cite{Ronchi1996, Taylor1997, Dennis2003}.
\begin{table}[ht]
\renewcommand{\arraystretch}{1.5}
\begin{center}
\caption{Cube-Center Locations for $S^2\times S^1$
\label{t:TableII}}
\begin{tabular}{l|l|l}
\hline\hline
$\vec c_1 =( 0, -L,  0)$ & $\vec c_3 =( 0, L,  0)$ & $\vec c_5 =( L, 0,  0)$ \\
$\vec c_2 =( 0, 0,  0)$ & $\vec c_4 =( 0, 2L,  0)$ & $\vec c_6 =( -L, 0,  0)$ \\
\hline\hline
 \end{tabular}
\end{center}
\end{table}
\begin{table}[t|]
\renewcommand{\arraystretch}{1.5}
\begin{center}
\caption{Cube Face Identifications,
$\partial_\alpha{\cal B}_A \leftrightarrow \partial_\beta {\cal B}_B$,
 and rotation matrices, ${\mathbf C}^{A\alpha}_{B\beta}$, for 
the interface maps in $S^2\times S^1$.
\label{t:TableIII}}
\begin{tabular}{ccc||ccc}
\hline\hline
$\partial_\alpha{\cal B}_A \leftrightarrow \partial_\beta {\cal B}_B$
&${\mathbf C}^{A\alpha}_{B\beta}$
&${\mathbf C}^{B\beta}_{A\alpha}$
&$\partial_\alpha{\cal B}_A \leftrightarrow \partial_\beta {\cal B}_B$
&${\mathbf C}^{A\alpha}_{B\beta}$
&${\mathbf C}^{B\beta}_{A\alpha}$
\\
\hline
$\partial_{+z}{\cal B}_1 \leftrightarrow \partial_{-z} {\cal B}_1$
& ${\mathbf I}$ & ${\mathbf I}$
& $\partial_{+y}{\cal B}_1 \leftrightarrow \partial_{-y} {\cal B}_2$
& ${\mathbf I}$ & ${\mathbf I}$\\ 
$\partial_{-y}{\cal B}_1 \leftrightarrow \partial_{+y} {\cal B}_4$
& ${\mathbf I}$ & ${\mathbf I}$
& $\partial_{+x}{\cal B}_1 \leftrightarrow \partial_{-y} {\cal B}_5$
& ${\mathbf R}_{+z}$ & ${\mathbf R}_{-z}$\\ 
$\partial_{-x}{\cal B}_1 \leftrightarrow \partial_{-y} {\cal B}_6$
& ${\mathbf R}_{-z}$ & ${\mathbf R}_{+z}$
& $\partial_{+z}{\cal B}_2 \leftrightarrow \partial_{-z} {\cal B}_2$
& ${\mathbf I}$ & ${\mathbf I}$\\
$\partial_{+y}{\cal B}_2 \leftrightarrow \partial_{-y} {\cal B}_3$
& ${\mathbf I}$ & ${\mathbf I}$
& $\partial_{+x}{\cal B}_2 \leftrightarrow \partial_{-x} {\cal B}_5$
& ${\mathbf I}$ & ${\mathbf I}$\\
$\partial_{-x}{\cal B}_2 \leftrightarrow \partial_{+x} {\cal B}_6$
& ${\mathbf I}$ & ${\mathbf I}$
& $\partial_{+z}{\cal B}_3 \leftrightarrow \partial_{-z} {\cal B}_3$
& ${\mathbf I}$ & ${\mathbf I}$\\
$\partial_{+y}{\cal B}_3 \leftrightarrow \partial_{-y} {\cal B}_4$
& ${\mathbf I}$ & ${\mathbf I}$
& $\partial_{+x}{\cal B}_3 \leftrightarrow \partial_{+y} {\cal B}_5$
& ${\mathbf R}_{-z}$ & ${\mathbf R}_{+z}$\\
$\partial_{-x}{\cal B}_3 \leftrightarrow \partial_{+y} {\cal B}_6$
& ${\mathbf R}_{+z}$ & ${\mathbf R}_{-z}$
& $\partial_{+z}{\cal B}_4 \leftrightarrow \partial_{-z} {\cal B}_4$
& ${\mathbf I}$ & ${\mathbf I}$\\
$\partial_{+x}{\cal B}_4 \leftrightarrow \partial_{+x} {\cal B}_5$
& ${\mathbf R}_{+z}^2$ & ${\mathbf R}_{+z}^2$
& $\partial_{-x}{\cal B}_4 \leftrightarrow \partial_{-x} {\cal B}_6$
& ${\mathbf R}_{+z}^2$ & ${\mathbf R}_{+z}^2$\\
$\partial_{+z}{\cal B}_5 \leftrightarrow \partial_{-z} {\cal B}_5$
& ${\mathbf I}$ & ${\mathbf I}$
& $\partial_{+z}{\cal B}_6 \leftrightarrow \partial_{-z} {\cal B}_6$
& ${\mathbf I}$ & ${\mathbf I}$
\\
\hline\hline
 \end{tabular}
\end{center}
\end{table}

It is useful to discuss the method used to construct the
``cubed-sphere'' representation of $S^2$ in some detail here, since
this method is used in \ref{s:ExampleS3} as the model for constructing
a new representation of $S^3$.  Let $\{\bar x, \bar y, \bar z\}$
denote Cartesian coordinates in an $R^3$, and let $\bar x^2 + \bar y^2
+\bar z^2= r^2$ denote a two-sphere $S^2$ of radius $r$.  It is
useful for some purposes to identify points on this $S^2$ using
standard angular coordinates $\theta$ and $\varphi$:
\begin{eqnarray}
\bar x & = & r \sin\theta\cos\varphi,\label{e:xbar}\\
\bar y & = & r \sin\theta\sin\varphi,\\
\bar z & = & r \cos\theta.\label{e:wbar}
\end{eqnarray}
Now consider a cube $\bar{\cal B}$ centered at the origin, of size
$L=2r/\sqrt{3}$ (which just fits inside the sphere), whose orientation
is aligned with the $\{\bar x, \bar y, \bar z\}$ axes.  Let
$\partial_{\bar\alpha}{\cal \bar B}$ represent the six faces of this
cube, with $\bar\alpha=\pm \bar x$, etc., labeling the various faces.
The images of these six faces can be arranged in a plane, like the
$\alpha=+z$ faces of the cubes shown in Fig.~\ref{f:s2_x_s1}.  The
goal here is to construct a representation of $S^2\times S^1$, so it
will also be useful to make a correspondence between these cube faces
$\partial_{\bar\alpha}{\cal \bar B}$ with the cubes shown in
Fig.~\ref{f:s2_x_s1}.  Table~\ref{t:TableIV} gives the relationship
between the cube-face identifiers $\bar \alpha=\pm\bar x$, etc.  and
the cubic region labels $\scriptstyle{A}=1,2, ..., 6$ shown in
Fig.~\ref{f:s2_x_s1}.

Points on each of the cube-faces, $\partial_{\bar\alpha}{\cal \bar
  B}$, can be identified by their local Cartesian coordinates.  For
example, points on the $\bar \alpha =+\bar z$ face, i.e. the
${\scriptstyle A}=2$ face in Fig.~\ref{f:s2_x_s1}, can be identified
by the coordinates $\{\bar x, \bar y\}$.  It is also useful to
introduce scaled local Cartesian coordinates, $\{X_A,Y_A\}$ to
represent the points on these faces.  For the $\bar\alpha=+\bar z$
face for example, it is useful to set $\{X_2,Y_2\}=\{\bar x/\bar
z,\bar y/\bar z\}$.  Each coordinate has been divided by $\bar z$,
which is constant on this face, to ensure that the scaled coordinates
$\{X_2,Y_2\}$ are confined to the ranges, $-1\leq X_2\leq 1$ and
$-1\leq Y_2\leq 1$.  Similar definitions are made on the other faces,
cf. Table~\ref{t:TableIV}, that ensure the $X_A$ and $Y_A$ are all
oriented the same way as in Fig.~\ref{f:s2_x_s1}, and all satisfy
$-1\leq X_A\leq 1$ and $-1\leq Y_A\leq 1$.  Using
Eqs.~(\ref{e:xbar})--(\ref{e:wbar}), this construction provides a
natural identification between points on the original sphere, labeled
by their angular coordinates $\{\theta, \varphi\}$, and the
Cartesian cube-face coordinates $\{X_A,Y_A\}$ via the equations
summarized in Tables~\ref{t:TableIV} and \ref{t:TableV}.
\begin{table}[ht]
\renewcommand{\arraystretch}{1.5}
\begin{center}
\caption{Cubed-Sphere Representation of $S^2$:
Angular to Cartesian Coordinate Map.
\label{t:TableIV}}
\begin{tabular}{c|c|rcl|rcl}
\hline\hline
${\scriptstyle A}$ &$\bar\alpha$
&&$X_A$ &&&$Y_A$ &  \\
\hline
1 & $-\bar y$ 
& $-\frac{\bar x}{\bar y}$ &=&$-\cot\varphi$ 
& $-\frac{\bar z}{\bar y}$ &=&$-\cot\theta\csc\varphi$ \\
2 & $+\bar z$ 
& $\frac{\bar x}{\bar z}$ &=&$\tan\theta\cos\varphi$ 
& $\frac{\bar y}{\bar z}$ &=&$\tan\theta\sin\varphi$ \\
3 & $+\bar y$ 
& $\frac{\bar x}{\bar y}$ &=&$\cot\varphi$ 
& $-\frac{\bar z}{\bar y}$ &=&$-\cot\theta\csc\varphi$ \\
4 & $-\bar z$ 
& $-\frac{\bar x}{\bar z}$ &=&$-\tan\theta\cos\varphi$ 
& $\frac{\bar y}{\bar z}$ &=&$\tan\theta\sin\varphi$ \\
5 & $+\bar x$ 
& $-\frac{\bar z}{\bar x}$ &=&$-\cot\theta\sec\varphi$ 
& $\frac{\bar y}{\bar x}$ &=&$\tan\varphi$ \\
6 & $-\bar x$ 
& $-\frac{\bar z}{\bar x}$ &=& $-\cot\theta\sec\varphi$ 
& $-\frac{\bar y}{\bar x}$ &=&$-\tan\varphi$ \\
\hline\hline
 \end{tabular}
\end{center}
\end{table}
\begin{table*}[ht]
\renewcommand{\arraystretch}{1.7}
\begin{center}
\caption{ Cartesian to Angular Coordinate Map for the Cubed-Sphere
  Representation of $S^2$.  The range of the local Cartesian
  coordinate $X_A$ is $-1\leq X_A\leq 1$, and the range of $\theta$ is
  $0\leq\theta\leq\pi$ in these expressions.  The ranges of $\varphi$
  for different values of $Y_A$ are specified in the table.
\label{t:TableV}}
\begin{tabular}{c|c|c|c|c}
\hline\hline
${\scriptstyle A}$
&$Y_A$-range 
& $\cos \varphi$ &$\varphi$-range  
&$\cos\theta$   \\
\hline 
1 
& $-1\leq Y_1 \leq 1$
& $X_1/\sqrt{1+X_{1}^2}$ 
& $\frac{7\pi}{4}\geq\varphi\geq\frac{5\pi}{4}$ 
& $Y_1/{\sqrt{1+X_{1}^2+Y_{1}^2}} $
\\

2 
&  $1\geq Y_{2}\geq 0$
& $X_2/\sqrt{X_{2}^2+Y_{2}^2}$ 
& $\pi\geq\varphi\geq 0$
& $1/{\sqrt{1+X_{2}^2+Y_{2}^2}} $
\\
2 
&  $-1\leq Y_{2}< 0$ 
& $X_2/\sqrt{X_{2}^2+Y_{2}^2}$ 
& $2\pi>\varphi\geq {\pi}$
& $1/{\sqrt{1+X_{2}^2+Y_{2}^2}} $
\\
3 
& $-1\leq Y_3 \leq 1$
& $X_3/\sqrt{1+X_{3}^2}$ 
& $\frac{3\pi}{4}\geq\varphi\geq\frac{\pi}{4}$ 
& $-Y_3/{\sqrt{1+X_{3}^2+Y_{3}^2}} $
\\

4 
&  $1\geq Y_{4}> 0$ 
& $X_4/\sqrt{X_{4}^2+Y_{4}^2}$ 
& $2\pi>\varphi\geq \pi$
& $-1/{\sqrt{1+X_{4}^2+Y_{4}^2}} $
\\

4 
&  $-1\leq Y_{4}\leq 0$ 
& $X_4/\sqrt{X_{4}^2+Y_{4}^2}$ 
& $\pi\geq\varphi\geq 0$
& $-1/{\sqrt{1+X_{4}^2+Y_{4}^2}} $
\\

5
& $-1\leq Y_{5}< 0$
& $1/\sqrt{1+Y_{5}^2}$ 
& ${2\pi}>\varphi\geq\frac{7\pi}{4}$ 
& $-X_5/{\sqrt{1+X_{5}^2+Y_{5}^2}}$
\\

5
& $1\geq Y_{5}\geq 0$
& $1/\sqrt{1+Y_{5}^2}$ 
& $\frac{\pi}{4}\geq\varphi\geq 0$ 
& $-X_5/{\sqrt{1+X_{5}^2+Y_{5}^2}}$
\\

6
& $-1\leq Y_{6}< 0$
& $-1/\sqrt{1+Y_{6}^2}$ 
& $\frac{5\pi}{4}\geq\varphi>\pi$ 
& $X_6/{\sqrt{1+X_{6}^2+Y_{6}^2}}$
\\

6
& $1\geq Y_{6}\geq 0$
& $-1/\sqrt{1+Y_{6}^2}$ 
& $\pi\geq\varphi\geq\frac{3\pi}{4}$ 
& $X_6/{\sqrt{1+X_{6}^2+Y_{6}^2}}$
\\

\hline\hline
 \end{tabular}
\end{center}
\end{table*}

 The $\{X_A,Y_A\}$ defined in this way are local Cartesian
 coordinates.  These could be converted to global coordinates by
 adding in the appropriate offset for each face:
 $x_A^x=c^x_A+\frac{1}{2}L X_A$ and $x_A^y=c^y_A+\frac{1}{2}L Y_A$.
 Alternatively, the angles $\tan^{-1}X_A$ and $\tan^{-1}Y_A$ could be
 used as local ``Cartesian'' coordinates on these cube faces.  These
 angle-based Cartesian coordinates have the advantage of giving a more
 uniform mapping of the Euclidean plane onto the image of the cube
 face on the sphere, so they are the preferred choice for numerical
 work.  Global Cartesian coordinates constructed from these
 angle-based coordinates are defined by
\begin{eqnarray}
x_A^x=c_A^x+\frac{2L}{\pi}\tan^{-1}X_A,\label{e:X_A}\\
x_A^y=c_A^y+\frac{2L}{\pi}\tan^{-1}Y_A,\label{e:Y_A}
\end{eqnarray}
where $X_A$ and $Y_A$ are functions of the standard angular
coordinates $\theta$ and $\varphi$ by the expressions given in
Table~\ref{t:TableIV}.  

For representations of $S^2\times S^1$, an appropriate coordinate is
also needed for the periodically identified $z$ direction in
Fig.~\ref{f:s2_x_s1}.  Introduce an angle $\chi$, whose range is
$-\pi\leq \chi\leq\pi$, that labels the points in the $S^1$ subspace.
Then define the global Cartesian coordinate associated with this
direction as
\begin{eqnarray}
x_A^z=c_A^z+\frac{L}{2\pi}\chi,\label{e:Z_A}
\end{eqnarray}

The standard constant-curvature ``round'' metric on $S^2\times S^1$ is
smooth, and it is therefore an acceptable choice for the reference
metric to define the differential structure on this manifold.  The
simplest representation of this round metric uses the angular
coordinates $\theta$, $\varphi$, and $\chi$:
\begin{eqnarray}
ds^2= 
R_2^2(d\theta^2 + \sin^2\theta d\varphi^2)
+R_1^2d\chi^2,
\label{e:S2xS1RoundMetric}
\end{eqnarray}
where $R_2$ and $R_1$ are constants that specify the radii of the
$S^2$ and $S^1$ parts of the geometry respectively.  Using the
transformations given in Eqs.~(\ref{e:X_A})--(\ref{e:Z_A}) and
Table~\ref{t:TableIV}, a straightforward (but lengthy) calculation
gives the global multi-cube Cartesian-coordinate representation of
this metric on $S^2\times S^1$:
\begin{eqnarray}
\!\!\!\!\!
ds^2 &=& 
\left(\frac{\pi R_2}{2L}\right)^2 
\frac{(1+X_A^2)(1+Y_A^2)}{(1+X_A^2+Y_A^2)^2}
\Bigl[(1+X_A^2)(dx_A^x)^2
-2X_AY_A dx_A^xdx_A^y
+(1+Y_A^2)(dx_A^y)^2\Bigr]\nonumber\\
&&+
\left(\frac{2\pi R_1}{L}\right)^2(dx_A^z)^2.
\label{e:S2xS1CartesianRoundMetric}
\end{eqnarray}
The $X_A$ and $Y_A$ that appear in this expression are thought of as
the functions of the Cartesian coordinates obtained by inverting the
expressions given in Eqs.~(\ref{e:X_A}) and (\ref{e:Y_A}):
\begin{eqnarray}
X_A= \tan\left[\frac{\pi(x_A^x - c_A^x)}{2L}\right],\\
Y_A= \tan\left[\frac{\pi(x_A^y - c_A^y)}{2L}\right].
\end{eqnarray}
The functions $X_A$ and $Y_A$ depend on the location of a particular
coordinate region through the parameters $c_A^x$ and $c_A^y$.
However, beyond this dependence the multi-cube coordinate
representation of the $S^2\times S^1$ round metric given in
Eq.~(\ref{e:S2xS1CartesianRoundMetric}) is the same in each of the six
coordinate regions ${\cal B}_A$.

These multi-cube Cartesian coordinates $\{x_A,y_A,z_A\}$ turn out to
be harmonic with respect to the round metric on $S^2\times S^1$, i.e.,
each coordinate is a solution (locally within each cubic-region, not
globally across the interface boundaries) to the covariant Laplace
equation, $0=\nabla^i_A\nabla_{Ai}\, x_A= \nabla^i_A\nabla_{Ai}\,
y_A=\nabla^i_A\nabla_{Ai}\, z_A$, where $\nabla_{Ai}$ is the covariant
derivative associated with the $S^2\times S^1$ metric in region
${\scriptstyle A}$.  These conditions are equivalent to
$0=\partial_{Ai}\left(\sqrt{g_A}\, g^{ij}_A\right)$ where $g_A=\det
g_{Aij}$ and $g^{ij}_A$ is the inverse of the metric $g_{Aij}$
expressed in terms of the multi-cube Cartesian coordinates in region
${\scriptstyle A}$.

\subsection{Multi-Cube Representation of $S^3$}
\label{s:ExampleS3} 

\begin{figure}[htb|]
\centerline{\includegraphics[width=2.5in]{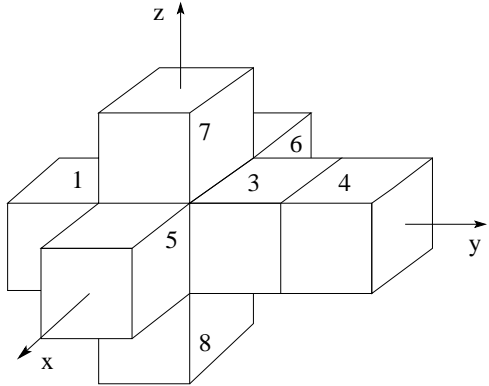}}
\caption{\label{f:s3} The three-manifold $S^3$ can be represented
  using the eight cubic regions illustrated here.  Cubic region ${\cal
    B}_2$, centered at the origin $\vec c_2=(0,0,0)$ is hidden between
  ${\cal B}_7$ and ${\cal B}_8$ in this figure.  The outer faces of
  these cubes are identified using the maps described in
  Table~\ref{t:TableVII}. }
\end{figure}

The locations of the eight cubic regions used to construct this
representation of $S^3$ are illustrated in Fig.~\ref{f:s3}.  The
values of the cube-center location vectors $\vec c_A$ for this
configuration is summarized in Table~\ref{t:TableVI}.  The inner
faces of the touching cubes in Fig.~\ref{f:s3} are assumed to be
connected by identity maps.  The outer faces of these eight cubic
regions are identified using the maps described in
Table~\ref{t:TableVII}.  This ``cubed-sphere'' representation of $S^3$
is a natural three-dimensional generalization of the two-dimensional
cubed-sphere representation of $S^2$ described in
\ref{s:ExampleS2xS1}.  It is constructed by inserting a
four-dimensional cube into a three-dimensional sphere $S^3$ in $R^4$,
and then identifying points on the faces of the four-cube with the
points on the three-sphere that are connected by rays extending
outward from the origin.
\begin{table}[ht]
\renewcommand{\arraystretch}{1.5}
\begin{center}
\caption{Cube-Center Locations for $S^3$
\label{t:TableVI}}
\begin{tabular}{l|l|l|l}
\hline\hline 
$\vec c_1 =( 0, -L, 0)$ & $\vec c_3 =( 0, L, 0)$ & $\vec c_5 =( L, 0, 0)$ 
& $\vec c_7 =( 0, 0, L)$ \\ 
$\vec c_2 =( 0, 0, 0)$ & $\vec c_4 =( 0, 2L, 0)$ & $\vec c_6 =( -L, 0, 0)$ 
& $\vec c_8 =( 0, 0, -L)$ \\ 
\hline\hline
 \end{tabular}
\end{center}
\end{table}
\begin{table}[htb|]
\renewcommand{\arraystretch}{1.5}
\begin{center}
\caption{Cubic Region Face Identifications, $\partial_\alpha{\cal B}_A
  \leftrightarrow \partial_\beta {\cal B}_B$ , and
  rotation matrices, ${\cal C}^{A\alpha}_{B\beta}$, for the 
interface maps in $S^3$.
\label{t:TableVII}}
\begin{tabular}{c|c|c||c|c|c}
\hline\hline
$\partial_\alpha{\cal B}_A \leftrightarrow \partial_\beta {\cal B}_B$
&${\mathbf C}^{A\alpha}_{B\beta}$
&${\mathbf C}^{B\beta}_{A\alpha}$
&$\partial_\alpha{\cal B}_A \leftrightarrow \partial_\beta {\cal B}_B$
&${\mathbf C}^{A\alpha}_{B\beta}$
&${\mathbf C}^{B\beta}_{A\alpha}$
\\
\hline
$\partial_{+y}{\cal B}_1 \leftrightarrow \partial_{-y} {\cal B}_2$
& ${\mathbf I}$ & ${\mathbf I}$ 
& $\partial_{-y}{\cal B}_1 \leftrightarrow \partial_{+y} {\cal B}_4$
& ${\mathbf I}$ & ${\mathbf I}$\\
$\partial_{+x}{\cal B}_1 \leftrightarrow \partial_{-y} {\cal B}_5$
& ${\mathbf R}_{+z}$ & ${\mathbf R}_{-z}$ 
& $\partial_{-x}{\cal B}_1 \leftrightarrow \partial_{-y} {\cal B}_6$
& ${\mathbf R}_{-z}$ & ${\mathbf R}_{+z}$\\
$\partial_{+z}{\cal B}_1 \leftrightarrow \partial_{-y} {\cal B}_7$
& ${\mathbf R}_{-x}$ & ${\mathbf R}_{+x}$ 
& $\partial_{-z}{\cal B}_1 \leftrightarrow \partial_{-y} {\cal B}_8$
& ${\mathbf R}_{+x}$ & ${\mathbf R}_{-x}$\\
$\partial_{+y}{\cal B}_2 \leftrightarrow \partial_{-y} {\cal B}_3$
& ${\mathbf I}$ & ${\mathbf I}$
& $\partial_{+x}{\cal B}_2 \leftrightarrow \partial_{-x} {\cal B}_5$
& ${\mathbf I}$ & ${\mathbf I}$\\
$\partial_{-x}{\cal B}_2 \leftrightarrow \partial_{+x} {\cal B}_6$
& ${\mathbf I}$ & ${\mathbf I}$
& $\partial_{+z}{\cal B}_2 \leftrightarrow \partial_{-z} {\cal B}_7$
& ${\mathbf I}$ & ${\mathbf I}$\\
$\partial_{-z}{\cal B}_2 \leftrightarrow \partial_{+z} {\cal B}_8$
& ${\mathbf I}$ & ${\mathbf I}$
& $\partial_{+y}{\cal B}_3 \leftrightarrow \partial_{-y} {\cal B}_4$
& ${\mathbf I}$ & ${\mathbf I}$\\
$\partial_{+x}{\cal B}_3 \leftrightarrow \partial_{+y} {\cal B}_5$
& ${\mathbf R}_{-z}$ & ${\mathbf R}_{+z}$ 
& $\partial_{-x}{\cal B}_3 \leftrightarrow \partial_{+y} {\cal B}_6$
& ${\mathbf R}_{+z}$ & ${\mathbf R}_{-z}$\\
$\partial_{+z}{\cal B}_3 \leftrightarrow \partial_{+y} {\cal B}_7$
& ${\mathbf R}_{+x}$ & ${\mathbf R}_{-x}$
& $\partial_{-z}{\cal B}_3 \leftrightarrow \partial_{+y} {\cal B}_8$
& ${\mathbf R}_{-x}$ & ${\mathbf R}_{+x}$\\
$\partial_{+x}{\cal B}_4 \leftrightarrow \partial_{+x} {\cal B}_5$
& ${\mathbf R}_{+z}^2$ & ${\mathbf R}_{+z}^2$
& $\partial_{-x}{\cal B}_4 \leftrightarrow \partial_{-x} {\cal B}_6$
& ${\mathbf R}_{+z}^2$ & ${\mathbf R}_{+z}^2$\\
$\partial_{+z}{\cal B}_4 \leftrightarrow \partial_{+z} {\cal B}_7$
& ${\mathbf R}_{+x}^2$ & ${\mathbf R}_{+x}^2$
& $\partial_{-z}{\cal B}_4 \leftrightarrow \partial_{-z} {\cal B}_8$
& ${\mathbf R}_{+x}^2$ & ${\mathbf R}_{+x}^2$\\
$\partial_{+z}{\cal B}_5 \leftrightarrow \partial_{+x} {\cal B}_7$
& ${\mathbf R}_{-y}$ & ${\mathbf R}_{+y}$
& $\partial_{-z}{\cal B}_5 \leftrightarrow \partial_{+x} {\cal B}_8$
& ${\mathbf R}_{+y}$ & ${\mathbf R}_{-y}$\\
$\partial_{+z}{\cal B}_6 \leftrightarrow \partial_{-x} {\cal B}_7$
& ${\mathbf R}_{+y}$ & ${\mathbf R}_{-y}$
& $\partial_{-z}{\cal B}_6 \leftrightarrow \partial_{-x} {\cal B}_8$
& ${\mathbf R}_{-y}$ & ${\mathbf R}_{+y}$ \\
\hline\hline
 \end{tabular}
\end{center}
\end{table}

It is appropriate to discuss this ``cubed-sphere'' representation of
$S^3$ in some detail, since it does not appear to have been used or
described in the literature before.  Let $\{\bar x, \bar y, \bar
z,\bar w\}$ denote Cartesian coordinates in $R^4$, and let $\bar
x^2 + \bar y^2 +\bar z^2+\bar w^2= r^2$ denote a three-sphere, $S^3$,
of radius $r$.  It is often useful to identify points in $S^3$ using
the angular coordinates $\chi$, $\theta$ and $\varphi$:
\begin{eqnarray}
\bar x = r \sin\chi\sin\theta\cos\varphi,
\label{e:S3barXDef}\\
\bar y = r \sin\chi\sin\theta\sin\varphi,\\
\bar z = r \sin\chi\cos\theta,\\
\bar w = r \cos\chi.
\label{e:S3barWDef}
\end{eqnarray}
Now consider a four-cube centered at the origin, of size $L=r$ (which
just fits inside the three-sphere), whose orientation is aligned with
the $\{\bar x, \bar y, \bar z,\bar w\}$ axes.  Let
$\partial_{\bar\alpha}{\cal \bar B}$ denote the eight faces of this
four-cube (each of which is a three-cube) labeled by the index
$\bar\alpha=\pm \bar x$, etc.  Arrange the images of these eight
three-cubes in $R^3$ at the locations given in Table~\ref{t:TableVI},
as shown in Fig.~\ref{f:s3}.  Table~\ref{t:TableVIII} gives the
relationship between the four-cube face identifiers $\bar
\alpha=\pm\bar x$, etc.  and the three-cube region identifiers
$\scriptstyle{A}=1,2, ..., 8$ shown in Fig.~\ref{f:s3}.

Points on each of the four-cube faces, $\partial_{\bar\alpha}{\cal
  \bar B}$, can be identified by their local Cartesian coordinates.
For example, points on the $\bar \alpha =+\bar w$ face, i.e. the
${\scriptstyle A}=2$ region in Fig.~\ref{f:s3}, can be identified by
the coordinates $\{\bar x, \bar y,\bar z\}$.  It is convenient to
introduce scaled local Cartesian coordinates, $\{X_A,Y_A,Z_A\}$ to
represent the points on these faces.  For the $\bar\alpha=+\bar w$
face for example, set $\{X_2,Y_2,Z_2\}=\{\bar x/\bar w,\bar y/\bar
w,\bar z/\bar w\}$.  Each coordinate has been divided by $\bar w$,
which is constant on this face, to ensure that the scaled coordinates
$\{X_2,Y_2,Z_2\}$ are confined to the ranges, $-1\leq X_2\leq 1$,
$-1\leq Y_2\leq 1$, and $-1\leq Z_2\leq 1$.  Similar definitions are
made on the other faces, cf. Table~\ref{t:TableVIII}, that ensure the
$X_A$, $Y_A$, and $Z_A$ are all oriented the same way as in
Fig.~\ref{f:s3}, and all satisfy $-1\leq X_A\leq 1$, $-1\leq Y_A\leq
1$, and $-1\leq Z_A\leq 1$.  Using
Eqs.~(\ref{e:S3barXDef})--(\ref{e:S3barWDef}), this construction
provides a natural identification between points on the original
three-sphere, labeled by their angular coordinates $\{\chi,\theta,
\varphi\}$, and the local Cartesian coordinates $\{X_A,Y_A,Z_A\}$ on
each four-cube face via the equations summarized in
Tables~\ref{t:TableVIII} and \ref{t:TableIX}.
\begin{table*}[ht]
\renewcommand{\arraystretch}{1.5}
\begin{center}
\caption{Cubed-Sphere Representation of
  $S^3$.
\label{t:TableVIII}}
\begin{tabular}{c|c|c|c|c}
\hline\hline
${\scriptstyle A}$ &$\bar\alpha$
&$X_A$ &$Y_A$ &$Z_a$   \\
\hline
1 & $-\bar y$ 
& $-\frac{\bar x}{\bar y}$ = $-\cot\varphi$ 
& $-\frac{\bar w}{\bar y}$= $-\cot\chi\csc\theta\csc\varphi$
& $-\frac{\bar z}{\bar y}$ = $-\cot\theta\csc\varphi$ \\
2 & $+\bar w$ 
& $\frac{\bar x}{\bar w}$ = $\tan\chi\sin\theta\cos\varphi$
& $\frac{\bar y}{\bar w}$ = $\tan\chi\sin\theta\sin\varphi$
& $\frac{\bar z}{\bar w}$ = $\tan\chi\cos\theta$ \\
3 & $+\bar y$ 
& $\frac{\bar x}{\bar y}$ = $\cot\varphi$
& $-\frac{\bar w}{\bar y}$ = $-\cot\chi\csc\theta\csc\varphi$
& $\frac{\bar z}{\bar y}$ = $\cot\theta\csc\varphi$ \\
4 & $-\bar w$ 
& $-\frac{\bar x}{\bar w}$ = $-\tan\chi\sin\theta\cos\varphi$
& $\frac{\bar y}{\bar w}$ = $\tan\chi\sin\theta\sin\varphi$
& $-\frac{\bar z}{\bar w}$ = $-\tan\chi\cos\theta$ \\
5 & $+\bar x$ 
& $-\frac{\bar w}{\bar x}$ = $-\cot\chi\csc\theta\sec\varphi$
& $\frac{\bar y}{\bar x}$ = $\tan\varphi$
& $\frac{\bar z}{\bar x}$ = $\cot\theta\sec\varphi$ \\
6 & $-\bar x$ 
& $-\frac{\bar w}{\bar x}$ = $-\cot\chi\csc\theta\sec\varphi$
& $-\frac{\bar y}{\bar x}$ = $-\tan\varphi$
& $-\frac{\bar z}{\bar x}$ = $-\cot\theta\sec\varphi$ \\
7 & $+\bar z$ 
& $\frac{\bar x}{\bar z}$ = $\tan\theta\cos\varphi$
& $\frac{\bar y}{\bar z}$ = $\tan\theta\sin\varphi$
& $-\frac{\bar w}{\bar z}$ = $-\cot\chi\sec\theta$ \\
8 & $-\bar z$ 
& $-\frac{\bar x}{\bar z}$ = $-\tan\theta\cos\varphi$  
& $-\frac{\bar y}{\bar z}$ = $-\tan\theta\sin\varphi$
& $-\frac{\bar w}{\bar z}$ = $-\cot\chi\sec\theta$ \\
\hline\hline
 \end{tabular}
\end{center}
\end{table*}

The $\{X_A,Y_A,Z_A\}$ defined using this cubed-sphere construction are
local Cartesian coordinates on each of the faces of the four-cube.
They could be converted to global coordinates by adding the
appropriate offset for each cube: $x_A^x=c^x_A+\frac{1}{2}L X_A$,
$x_A^y=c^y_A+\frac{1}{2}L Y_A$, and $x_A^z=c^z_A+\frac{1}{2}L Z_A$.
Alternatively, the angles $\tan^{-1}X_A$, $\tan^{-1}Y_A$, and
$\tan^{-1}Z_A$ also provide local Cartesian-like coordinates for these
cubes.  These angle-based Cartesian coordinates give a more uniform
mapping of Euclidean space onto the image of the four-cube face on the
three-sphere.  So as in the two-dimensional cubed-sphere case, these
angle-based Cartesian coordinates are the preferred choice for
numerical work on the multi-cube representation of $S^3$.  Global
multi-cube Cartesian coordinates constructed from these angle-based
coordinates are defined by
\begin{eqnarray}
x_A^x=c_A^x+\frac{2L}{\pi}\tan^{-1}X_A,\label{e:HX_A}\\
x_A^y=c_A^y+\frac{2L}{\pi}\tan^{-1}Y_A,\label{e:HY_A}\\
x_A^z=c_A^z+\frac{2L}{\pi}\tan^{-1}Z_A,\label{e:HZ_A}
\end{eqnarray}
where $X_A$, $Y_A$, and $Z_A$ are functions of the hyper-spherical
angular coordinates $\chi$, $\theta$ and $\varphi$ given by the expressions
in Tables~\ref{t:TableVIII} and \ref{t:TableIX}.
\begin{table*}[ht]
\renewcommand{\arraystretch}{1.7}
\begin{center}
\caption{ Cartesian to Angular Coordinate Map for the Cubed-Sphere
  Representation of $S^3$.  The range of the local Cartesian
  coordinate $X_A$ is $-1\leq X_A\leq 1$, the range of $Z_A$ is
  $-1\leq Z_A\leq 1$, the range of the angular coordinate $\theta$ is
  $0\leq \theta \leq \pi$, and the range of $\chi$ is $0\leq \chi\leq
  \pi$ in these expressions.  The ranges of $\varphi$ corresponding to
  different ranges of $Y_A$ are specified in the table.  The
  quantities $W_A \equiv \sqrt{1+X_A^2 + Y_A^2 + Z_A^2}$ are used to
  simplify the expressions for $\cos\chi$.
\label{t:TableIX}}

\begin{tabular}{c|c|c|c|c|c}
\hline\hline
${\scriptstyle A}$
&$Y_A$-range 
& $\cos \varphi$ &$\varphi$-range  
&$\cos\theta$ 
&$\cos\chi$ 
\\
\hline 

1 
& $-1\leq Y_1 \leq 1$
& $X_1/\sqrt{1+X_{1}^2}$ 
& $\frac{7\pi}{4}\geq\varphi\geq\frac{5\pi}{4}$
& $Z_1/{\sqrt{1+X_{1}^2+Z_{1}^2}} $
& $Y_1/W_1 $ 
\\

2 
& $1\geq Y_{2}\geq 0$
& ${X_2}/\sqrt{X_{2}^2+Y_2^2}$ 
& $\pi\geq\varphi\geq 0$
& $Z_2/{\sqrt{X_{2}^2+Y_2^2+Z_{2}^2}} $
& $1/W_2 $ 
\\

2 
& $-1\leq Y_{2}< 0$
& ${X_2}/\sqrt{X_{2}^2+Y_2^2}$ 
& $2\pi>\varphi\geq {\pi}$
& $Z_2/{\sqrt{X_{2}^2+Y_2^2+Z_{2}^2}} $
& $1/W_2 $ 
\\

3 
& $-1\leq Y_3 \leq 1$
& $X_3/\sqrt{1+X_{3}^2}$ 
& $\frac{3\pi}{4}\geq\varphi\geq\frac{\pi}{4}$
& $Z_3/{\sqrt{1+X_{3}^2+Z_{3}^2}} $
& $-Y_3/W_3 $ 
\\

4 
& $1\geq Y_{4}> 0$
& ${X_4}/\sqrt{X_{4}^2+Y_4^2}$ 
& $2\pi>\varphi\geq \pi$
& $Z_4/{\sqrt{X_{4}^2+Y_4^2+Z_{4}^2}} $
& $-1/W_4 $ 
\\

4 
& $-1\leq Y_{4}\leq 0$
& ${X_4}/\sqrt{X_{4}^2+Y_4^2}$ 
& $\pi\geq\varphi\geq 0$
& $Z_4/{\sqrt{X_{4}^2+Y_4^2+Z_{4}^2}} $
& $-1/W_4 $ 
\\

5
& $-1\leq Y_{5}< 0$
& $1/\sqrt{1+Y_5^2}$ 
& ${2\pi}>\varphi\geq\frac{7\pi}{4}$
& $Z_5/{\sqrt{1+Y_5^2+Z_{5}^2}} $
& $-X_5/W_5 $ 
\\

5
& $1\geq Y_{5}\geq 0$
& $1/\sqrt{1+Y_5^2}$ 
& $\frac{\pi}{4}\geq\varphi\geq 0$
& $Z_5/{\sqrt{1+Y_5^2+Z_{5}^2}} $
& $-X_5/W_5 $ 
\\

6
& $-1\leq Y_{6}< 0$
& $-1/\sqrt{1+Y_6^2}$ 
& $\frac{5\pi}{4}\geq\varphi>\pi$
& $Z_6/{\sqrt{1+Y_6^2+Z_{6}^2}} $
& $X_6/W_6 $ 
\\

6
& $1\geq Y_{6}\geq 0$
& $-1/\sqrt{1+Y_6^2}$ 
& $\pi\geq\varphi\geq\frac{3\pi}{4}$
& $Z_6/{\sqrt{1+Y_6^2+Z_{6}^2}} $
& $X_6/W_6 $ 
\\

7
& $1\geq Y_{7}\geq 0$
& ${X_7}/\sqrt{X_{7}^2+Y_7^2}$ 
&  $\pi\geq\varphi\geq 0$
& $1/{\sqrt{1+X_7^2+Y_7^2}} $
& $-Z_7/W_7 $ 
\\

7
& $-1\leq Y_{7}< 0$
& ${X_7}/\sqrt{X_{7}^2+Y_7^2}$ 
& $2\pi>\varphi\geq {\pi}$
& $1/{\sqrt{1+X_7^2+Y_7^2}} $
& $-Z_7/W_7 $ 
\\

8
& $1\geq Y_{8}\geq 0$
& ${X_8}/\sqrt{X_{8}^2+Y_8^2}$ 
& $\pi\geq\varphi\geq 0$
& $-1/{\sqrt{1+X_8^2+Y_8^2}} $
& $Z_8/W_8 $ 
\\

8
& $-1\leq Y_{8}< 0$
& ${X_8}/\sqrt{X_{8}^2+Y_8^2}$ 
& $2\pi>\varphi\geq {\pi}$
& $-1/{\sqrt{1+X_8^2+Y_8^2}} $
& $Z_8/W_8 $ 
\\

\hline\hline
 \end{tabular}
\end{center}
\end{table*}

The standard constant-curvature ``round'' metric on $S^3$ is smooth,
and it is therefore an acceptable choice for the reference metric to
define the differential structure on this manifold.  The simplest
representation of this round metric uses the angular coordinates
$\chi$, $\theta$, and $\varphi$:
\begin{eqnarray}
ds^2&=& R_3^2\left(d\chi^2 + \sin^2\!\chi\, d\theta^2
+ \sin^2\!\chi\sin^2\theta \,d\varphi^2\right),
\label{e:S^3RoundMetric}
\end{eqnarray}
where $R_3$ is a constant that specifies the radius of the $S^3$.
Using the transformations given in Eqs.~(\ref{e:HX_A})--(\ref{e:HZ_A})
and in Tables~\ref{t:TableVIII} and \ref{t:TableIX}, a straightforward
(but lengthy) calculation gives the global multi-cube
Cartesian-coordinate representation of this metric on $S^3$:
\begin{eqnarray}
ds^2&=& \left(\frac{\pi R_3}{2 L}\right)^2
\frac{(1+X_A^2)(1+Y_A^2)(1+Z_A^2)}{(1+X_A^2+Y_A^2+Z_A^2)^2}
\Biggl[\frac{(1+X_A^2)(1+Y_A^2+Z_A^2)}{(1+Y_A^2)(1+Z_A^2)}(dx^x_A)^2
-\frac{2X_AY_A}{1+Z_A^2}dx^x_Adx^y_A\nonumber\\
&&\qquad\qquad\qquad\qquad\qquad
+\frac{(1+Y_A^2)(1+X_A^2+Z_A^2)}{(1+X_A^2)(1+Z_A^2)}(dx^y_A)^2
-\frac{2X_AZ_A}{1+Y_A^2}dx^x_Adx^z_A\nonumber\\
&&\qquad\qquad\qquad\qquad\qquad
+\frac{(1+Z_A^2)(1+X_A^2+Y_A^2)}{(1+X_A^2)(1+Y_A^2)}(dx^z_A)^2
-\frac{2Y_AZ_A}{1+X_A^2}dx^y_Adx^z_A\Biggr].
\label{e:S^3CartesianRoundMetric}
\end{eqnarray}
The $X_A$, $Y_A$, and $Z_A$ that appear in
Eq.~(\ref{e:S^3CartesianRoundMetric}) are thought of as the functions
of the global multi-cube Cartesian coordinates obtained by inverting
the expressions given in Eqs.~(\ref{e:HX_A})--(\ref{e:HZ_A}):
\begin{eqnarray}
X_A&=& \tan\left[\frac{\pi(x_A^x - c_A^x)}{2L}\right],\\
Y_A&=& \tan\left[\frac{\pi(x_A^y - c_A^y)}{2L}\right],\\
Z_A&=& \tan\left[\frac{\pi(x_A^z - c_A^z)}{2L}\right].
\end{eqnarray}
The functions $X_A$, $Y_A$ and $Z_A$ depend on the location of a particular
coordinate region through the parameters $c_A^x$, $c_A^y$ and $c_A^z$.
However, beyond this dependence the multi-cube coordinate
representation of the $S^3$ round-sphere metric given in
Eq.~(\ref{e:S^3CartesianRoundMetric}) is the same in each of the eight
coordinate regions ${\cal B}_A$.

These multi-cube Cartesian coordinates $\{x_A,y_A,z_A\}$ turn out to
be harmonic with respect to the round metric on $S^3$, i.e.,
each coordinate is a solution (locally within each cubic-region, not
globally across the interface boundaries) to the covariant Laplace
equation, $0=\nabla^i_A\nabla_{Ai}\, x_A= \nabla^i_A\nabla_{Ai}\,
y_A=\nabla^i_A\nabla_{Ai}\, z_A$, where $\nabla_{Ai}$ is the covariant
derivative associated with the $S^3$ metric in region
${\scriptstyle A}$.  These conditions are equivalent to
$0=\partial_{Ai}\left(\sqrt{g_A} \,g^{ij}_A\right)$ where $g_A=\det
g_{Aij}$ and $g^{ij}_A$ is the inverse of the metric $g_{Aij}$
expressed in terms of the multi-cube Cartesian coordinates in region
${\scriptstyle A}$.

\section{Spherical Harmonics on $S^3$}
\label{s:HyperSphericalHarmonics}
This appendix derives expressions for the eigenfunctions of the
Laplace operator on the three-sphere $S^3$.  These eigenfunctions are
referred to here as three-sphere harmonics.  These functions are
defined as solutions of the equation
\begin{eqnarray}
\nabla^i\nabla_i Y = - \lambda Y,
\label{s:S3EigenvalueProblem}
\end{eqnarray}
where $\nabla_i$ is the covariant derivative operator on $S^3$, and
$\lambda$ is an eigenvalue.  These functions have been studied
previously by a number of authors~\cite{Lifshitz-Khalatnikov,
  Jantzen1978, Sandberg1978, Tomita1982}.  Here a slightly different
representation is introduced that allows these harmonics (of arbitrary
order) to be evaluated accurately in a straightforward way. Using the
angular coordinate representation of the round metric on $S^3$ from
Eq.~(\ref{e:S^3RoundMetric}), it is straightforward to write the
co-variant Laplace operator explicitly as
\begin{eqnarray}
\nabla^i\nabla_iY = 
\frac{\partial_\chi\left[\sin^2\chi\partial_\chi Y\right]}{R_3^2\sin^2\chi}
+\frac{\partial_\theta\left[\sin\theta\partial_\theta Y\right]}
{R_3^2\sin\theta\sin^2\chi}
+\frac{\partial_\varphi^{\,2}\,Y}{R_3^2\sin^2\theta\sin^2\chi}.
\end{eqnarray}
The eigenvalue problem, Eq.~(\ref{s:S3EigenvalueProblem}), can be
solved then by separation of variables.  The non-singular solutions to
this equation have the form:
\begin{eqnarray}
Y_{k\ell m}(\chi,\theta,\varphi)=\frac{N_{k\ell m}} {\sqrt{\sin\chi}}
Q^{\ell+\frac{1}{2}}_{k+\frac{1}{2}}(\cos\chi)P^m_\ell(\cos\theta)e^{i m\varphi},
\quad
\label{e:YklmSeparation}
\end{eqnarray}
where $P^\mu_\nu$ and $Q^\mu_\nu$ are the associated Legendre functions of
the first and second kind respectively.  The eigenvalue
associated with this $Y_{k\ell m}$ is
\begin{eqnarray}
\lambda = \frac{k(k+2)}{R_3^2}.
\end{eqnarray}
These functions are non-singular on $S^3$ only for integers $k$,
$\ell$ and $m$ satisfying
\begin{eqnarray}
&&k\geq 0,\\
&&k\geq\ell\geq 0,\\
&&\ell\geq m \geq -\ell.
\end{eqnarray}

The half-integer associated Legendre functions
$Q^{\ell+\frac{1}{2}}_{k+\frac{1}{2}}(x)$ with $x=\cos\chi$ are
non-singular for $-1\leq x \leq 1$, and can be evaluated re-cursively.
For fixed $\ell$, the functions with $k<\ell$ can be shown to vanish,
\begin{eqnarray}
Q^{\ell+\frac{1}{2}}_{k+\frac{1}{2}}(x)&=0,
\label{e:S3Qklgtkcase}
\end{eqnarray}
using \S3.4 Eq.~(13) in Ref.~\cite{Erdelyi1953v1}.  For $k=\ell$ a
similar argument using \S 3.6.1 Eq.~(14) in
Ref.~\cite{Erdelyi1953v1} gives
\begin{eqnarray}
Q^{\ell+\frac{1}{2}}_{\ell+\frac{1}{2}}(x)&=(-1)^{\ell+1}2^\ell\ell!\sqrt{\frac{\pi}{2}}
\left(1-x^2\right)^{\frac{\ell}{2}+\frac{1}{4}}.
\quad\end{eqnarray}
The functions with $k>\ell$ can be determined from these using the
recursion relation,
\begin{eqnarray}
(k-\ell+2)Q^{\ell+\frac{1}{2}}_{k+\frac{5}{2}}(x)=
2(k+2)\,x\,Q^{\ell+\frac{1}{2}}_{k+\frac{3}{2}}(x)
-(k+\ell+2)
Q^{\ell+\frac{1}{2}}_{k+\frac{1}{2}}(x),\qquad
\label{e:QlkRecursion}
\end{eqnarray}
from \S 3.8 Eq.~(12) in Ref.~\cite{Erdelyi1953v1}.
Evaluating Eq.~(\ref{e:QlkRecursion}) for $k=\ell-1$ gives
\begin{eqnarray}
Q^{\ell+\frac{1}{2}}_{\ell+\frac{3}{2}}(x)=2(\ell+1)\,x\,
Q^{\ell+\frac{1}{2}}_{\ell+\frac{1}{2}}(x),
\end{eqnarray}
using Eq.~(\ref{e:S3Qklgtkcase}).  The
$Q^{\ell+\frac{1}{2}}_{k+\frac{1}{2}}(x)$ with $k\geq\ell+2$ can then
be generated recursively using Eq.~(\ref{e:QlkRecursion}).  This
recursion relation is known to be a stable and accurate way to
generate the Legendre functions of the first kind, $P^m_\ell(x)$,
cf. Ref.~\cite{numrec_f}.  Our numerical tests indicate that it is also an
accurate way to generate the half-integer Legendre functions of the
second kind, $Q^{\ell+\frac{1}{2}}_{k+\frac{1}{2}}(x)$.

The orthogonality properties of the $Y_{k\ell m}(\chi,\theta,\varphi)$
are determined by the orthogonality properties of
$Q^{\ell+\frac{1}{2}}_{k+\frac{1}{2}}(\cos\chi)$,
$P^\ell_m(\cos\theta)$ and $e^{im\varphi}$.  The needed condition
for $Q^{\ell+\frac{1}{2}}_{k+\frac{1}{2}}$ can be obtained from the
associated Legendre differential equation,
\begin{eqnarray}
\!\!\!\!\!\!
0=\frac{d}{dx}\left[(1-x^2)\frac{dQ^\mu_\nu}{dx}\right]
\!+\!\left[\nu(\nu+1)-\frac{\mu^2}{1-x^2}\right]Q^\mu_\nu,
\end{eqnarray}
from which it follows that
\begin{eqnarray}
\frac{d}{dx}\left[(1-x^2)\left(Q^\mu_{\nu'}\frac{dQ^\mu_\nu}{dx}
-Q^\mu_{\nu}\frac{dQ^\mu_{\nu'}}{dx}\right)\right] =
(\nu'-\nu)(\nu+\nu'+1)Q^\mu_{\nu'}Q^\mu_{\nu}.
\label{e:QIdentity}
\end{eqnarray}
The half-integer associated Legendre functions are well behaved
in the interval $-1\leq x \leq 1$, therefore integrating
Eq.~(\ref{e:QIdentity}) over this interval gives
\begin{eqnarray}
0=(\nu'-\nu)(\nu+\nu'+1)\int_{-1}^1Q^\mu_{\nu'}(x)Q^\mu_{\nu}(x)dx.
\end{eqnarray}
It follows that the $Q^{\ell+\frac{1}{2}}_{k+\frac{1}{2}}(x)$
with $k\geq0$ and $\ell\geq0$ satisfy the orthogonality
condition:
\begin{eqnarray} 
M^2_{k\ell}\,\delta_{k'k} = 
\int_{-1}^1Q^{\ell+\frac{1}{2}}_{k'+\frac{1}{2}}(x)
Q^{\ell+\frac{1}{2}}_{k+\frac{1}{2}}(x)\,dx,
\end{eqnarray} 
where $M_{k\ell}$ is the numerical constant,
\begin{eqnarray}
M_{k\ell}^2 = \frac{\pi^2(k+\ell+1)!}{4(k+1)(k-\ell)!}.
\end{eqnarray}
The analogous orthogonality
relations for $P^\ell_m(\cos\theta)$ and $e^{im\varphi}$ are well known:
\begin{eqnarray}
N_{\ell m}^2 \delta_{\ell'\ell}=\int_{-1}^1P^m_{\ell'}(y)P^m_\ell(y)\,dy,\\
2\pi \delta_{m'm} =\int_0^{2\pi}e^{im'\varphi}e^{-im\varphi}d\varphi,
\end{eqnarray}
where
\begin{eqnarray}
N_{\ell m}^2=\frac{(\ell+m)!}{(\ell-m)!\left(\ell+\frac{1}{2}\right)}.
\end{eqnarray}
From these conditions then, it follows that by choosing the normalization
constants 
\begin{eqnarray}
N_{k\ell m} = \frac{1}{\sqrt{2\pi} M_{k\ell} N_{\ell m}},
\end{eqnarray}
the $Y_{k\ell m}$ satisfy
the following orthogonality conditions on $S^3$,
\begin{eqnarray}
\int Y_{k'\ell'm'} Y^*_{k\ell m} \sqrt{g}\,d^{\,3}x&=&
R_3^3 \int_0^\pi  d\chi\int_0^\pi d\theta\int_0^{2\pi}\!\!\!d\varphi
\,\sin^2\chi\,\sin\theta\, Y_{k'\ell'm'} Y^*_{k\ell m} 
  ,\nonumber\\
&=&\left[\frac{1}{M_{k\ell}^2}
\int_{-1}^1Q^{\ell'+\frac{1}{2}}_{k'+\frac{1}{2}}(x)Q^{\ell+\frac{1}{2}}_{k+\frac{1}{2}}(x)
\,dx\right]
\left[\frac{1}{N_{\ell m}^2}\int_{-1}^1P^m_{\ell'}(y)P^m_\ell(y)\,dy\right]
\nonumber\\
&&\times
\left[\frac{1}{2\pi}\int_0^{2\pi}e^{im'\varphi}e^{-im\varphi}d\varphi\right],\nonumber\\
&=&R_3^3\,\delta_{k'k}\delta_{\ell'\ell}\delta_{m'm}.
\end{eqnarray}

\vfill\eject
\bibliographystyle{model1-num-names} 
\bibliography{../References/References}

\end{document}